# Progress and Challenges for the Application of Machine Learning for Neglected Tropical Diseases

Chung Yuen Khew[1], Rahmad Akbar[2,C], and Norfarhan Mohd. Assaad[1,3C].

[1] Department of Applied Physics, Faculty of Science & Technology, Universiti Kebangsaan Malaysia, 43600 Bangi, MALAYSIA.
[2] Department of Immunology, University of Oslo and Oslo University Hospital, Norway
[3] Institute of Systems Biology (INBIOSIS), Universiti Kebangsaan Malaysia, 43600 Bangi, MALAYSIA.
[C] Correspondence

# Abstract

Neglected tropical diseases (NTDs) continue to affect the livelihood of individuals in countries in the Southeast Asia and Western Pacific region. These diseases have been long existing and have caused devastating health problems and economic decline to people in low- and middle-income (developing) countries. An estimated 1.7 billion of the world's population suffer one or more NTDs annually, this puts approximately one in five individuals at risk for NTDs. In addition to health and social impact, NTDs inflict significant financial burden to patients, close relatives, and are responsible for billions of dollars lost in revenue from reduced labor productivity in developing countries alone. There is an urgent need to better improve the control and eradication or elimination efforts towards NTDs. This can be achieved by utilizing machine learning tools to better the surveillance, prediction and detection program, and combat NTDs through the discovery of new therapeutics against these pathogens. This review surveys the current application of machine learning tools for NTDs and the challenges to elevate the state-of-the-art of NTDs surveillance, management, and treatment.

# Abbreviations

AAC: Amino Acid Composition; ACC: Scoring based on maximum accuracy; Adaboost; Adapative Boosting; AF2: AlphaFold2; AI: Artificial Intelligence; ALM: Antibody Language Model; ANN: Artificial Neural Network; AntiBERTa: Antibody-specific Bidirectional Encoder Representation from Transformers; ARACON: Asian Rabies Control Network; ASEAN: Association of Southeast Asian Nations; AUC: area under the ROC Curve; BERT: Bidirectional Encoder Representations from Transformers; C-ImmSim: C-language version of the IMMune system SIMulator; CASP14: Critical Assessment of Structure Prediction; CDRs: complementarity-determining regions; CNN: Convolutional neural network; CNS: Central Nervous System; Cryo-EM: cryogenic electron microscopy; CT: Computed Tomography; CTDC: Composition, transition, and distribution features; CTD: composition, transition and distribution; CTDC: CTD-composition; CTDD: CTD-distribution; CTDT: CTD-trancomposition; CV: Computer VisionCYD-TDV: chimeric yellow fever virus-DENV tetravalent dengue vaccine; DALYs: Disability-Adjusted Life Years; DBLP: Digital Bibliography and Library Project; DENV: Dengue Virus; DF: dengue fever; DHF: dengue haemorrhagic fever; DHS: Demographic and Health Surveys; DL: Deep Learning; DNA: deoxyribonucleic acid; DPC: dipeptide composition; DSS: dengue shock syndrome; DT: Decision Tree; EPG: eggs-per-gram; ECDC: The European Centre for Disease Prevention and Control; ESM-1: endothelial cell specific molecule-1; ESPEN: Expanded Special Project for the Elimination of NTDS; FFBPNN: Feed Forward Back Propagation Neural Network; GAAC: Grouped Amino Acid Composition; GAELF: Global Alliance to Eliminate Lymphatic Filariasis; GAHI: Global Atlas of Helminth Infections; GARC: Global Alliance for Rabies Control; GBD: Global Burden of Disease Study; GEMME: Global Epistatic Model Predicting Mutational Effects; GHDx: Global Health Data Exchange; GHE: Global Health Estimates; GHO: WHO Global Health Observatory; GLEAN: Global Leptospirosis Environment Action Network; GPELF: Global Programme to Eliminate Lymphatic Fliariasis; iAMAP-SCM: identification of antimalarial peptides with an interpretable scoring card method; IMGT: International IMmunoGeneTics information system; k-NN: k-Nearest Neighbour; LAD tree: Least Absolute Deviation Tree; LAMP: loop-mediated isothermal amplification; LF: Lymphatic Filariasis; LM: Language Models; LOPOV: leave-one-pathogen-out validation; LR: Logistic Regression; MAP: Malaria Atlas Project; MERACON:

Middle East, Eastern Europe, Central Asia and North Africa Rabies Control Network; MCC: Matthew's coefficient correlation; MDA: mass drug administration; ML: Machine Learning; MLP: Multilayer Perceptron; MMDP: Morbidity Management and Disability Prevention; MSA: multiple sequence alignments; MRI: Magnetic Resonance Imaging; N5CV: Nested 5-fold cross-validation; NB: Naïve Bayes; NCC: neurocysticercosis; NLP: Natural Language Processing; NMR: Nuclear Magnetic Resonance spectroscopy; NN: Neural Network; NTDs: Neglected Tropical Diseases; OAS: Observed Antibody Space; PARACON: Pan-African Rabies Control Network; PC: preventive chemotherapy; PCR: polymerase chain reaction; PET: Positron-emmission tomography; PLM: Protein Language Model; PLS: Partial Least Squares Regression; PPI: Protein-Protein Interaction; ProABC-2: Prediction of AntiBody contacts v2; QSAR: Quantitative structure-activity relationships; RF: Random Forest; RFE: Recursive Feature Elimination; RGN2: Recurrent Geometric Network; ROC: Receiver operating characteristics curve; RV: reverse vaccinology; SARS-Cov-2: Severe Acute Respiratory Syndrome Coronavirus 2; SEA: Southeast Asia; SIT; Sterile Insect Techniques; STHs: soil-transmitted helminthiasis; STRING: Search Tool for the Retrieval of Interacting Genes/Proteins; SVM: Support Vector Machine; TAS: Transmission Assessment Survey; TPC: tripeptide composition; WASH: Wash, Sanitation And Hygiene; Vaxign-ML: supervised machine learning reverse vaccinology model for improved prediction of bacterial protective antigens; VESPA: Variant Effect Score Prediction without Alignments; VTI: Voting Feature Interval; WHO: World Health Organization; XGB: Extreme Gradient Boosting; YLDs: Years of healthy life lost due to disability; YLLs: Years of life lost due to premature mortality.

# Introduction

## Neglected tropical diseases (NTDs)

Communicable diseases are illnesses caused by pathogens such as viruses, bacteria, parasitic worms, and fungi that can be contracted easily from contaminated surfaces, water, air droplets, air, bites from vector organisms, and direct contact with infected individuals. A recent Coronavirus disease (COVID-19) pandemic caused by the novel coronavirus Severe Acute Respiratory Syndrome Coronavirus 2 (SARS-CoV-2), with 623,470,447 confirmed cases of COVID-19, and 6,551,678 reported deaths, as of 20[th] October 2022 (https://covid19.who.int/), underlines the need for highly efficient approaches to manage, surveil, and develop new treatments for communicable diseases. Unlike the COVID-19 pandemic, a 'silent yet horrific endemic' caused by a diverse group of diseases affecting more than 149 countries and with more than 1.7 billion of infected individuals worldwide does not receive the same amount of attention (Jannin et al. 2017). The diverse group of diseases was coined by Peter Hotez and colleagues as 'neglected tropical diseases' (NTDs) as they do not receive the same urgency, in neither treatments nor surveillance, as other diseases such as AIDS, malaria, and tuberculosis (Winkler et al. 2018). The World Health Organization (WHO) first established a list of 17 "official" NTDs (World Health Organization 2010). Later in 2017, an addition of 3 disease conditions were made at the end of the 10[th] meeting of the Strategic and Technical Advisory Group for Neglected Tropical Diseases (World Health Organization 2017). A list of 20 NTDs recognized by WHO are shown in Table 1. Six of the listed NTDs (Dracunculiasis, Lymphatic filariasis, Onchocerciasis, Schistosomiasis, Soil-transmitted helminthiases, and Trachoma) can be controlled or eliminated through mass drug administrations (also called preventive chemotherapy), provision of safe water, sanitation and hygiene (WASH), innovative and intensified disease management, hygiene education programs have helped speed up efforts in eliminating these diseases (Hotez & Lo 2020;Zeynudin et al. 2022).

Table 1. List of NTDs recognized by WHO

| Category | Disease |
|---|---|
| i. Protozoan infections | 1. Chagas disease |
| | 2. Human African trypanosomiasis |
| | 3. Leishmaniasis |
| | |
| ii. Helminth infections | 4. Taeniasis/Cysticercosis |
| | 5. Dracunculiasis |
| | 6. Echinococcosis |
| | 7. Foodborne trematodes |
| | 8. Lymphatic filariasis |
| | 9. Onchocerciasis |
| | 10. Schistosomiasis |
| | 11. Soil-transmitted helminthiases |
| | - Ascariasis |
| | - Hookworm diseases |
| | - Trichuriasis |
| iii. Bacterial infections | 12. Buruli ulcer |
|  | 13. Leprosy |
|  | 14. Trachoma |
|  | 15. Yaws |
| iv. Viral infections | 16. Dengue and chikungunya fevers |
|  | 17. Rabies |
| v. Fungal Infections* | 18. Mycetoma, chromoblastomycosis, and other deep mycosis |
| vi. Ectoparasitic infections* | 19. Scabies, Myiasis |
| vii. Venom* | 20. Snakebite envenoming |

*Newly added diseases conditions into the NTD list prior to the outcome of 10[th] meeting of the Strategic and Technical Advisory Group for Neglected Tropical Diseases;
Source (https://www.who.int/health-topics/neglected-tropical-diseases#tab=tab_1)

## Disability-adjusted life year (DALY) impact of NTDs

High incidences of NTDs were commonly reported from tropical countries due to its optimal humidity and climate for the pathogens to thrive. Low- and middle-income countries in Africa and Asia that lack proper access to clean water and waste management greatly contribute to the spread of NTDs among women and children. To measure the extent of devastation caused by NTDs, the disability-adjusted life year (DALY; one DALY represents the loss of the equivalent of one year of full health) metric was introduced as a means to quantify the overall burden of disease borne by individuals (Mitra & Mawson 2017). DALYs for a disease or health condition are the sum of years of life lost due to premature mortality (YLLs) and years of healthy life lost due to disability (YLDs) due to prevalent cases of the disease or health condition in a population (Vinkeles Melchers et al. 2021). Based on the data collected by WHO, we were able to summarize the global burden for 14 of the 20 NTDs as estimated by DALYs in Table 2 below. Global burden for five of the highest estimated DALY burden NTDs are soil-transmitted helminthiases (STHs) (2.748 million years), rabies (2.635 million years), dengue (1.952 million years), schistosomiasis (1.628 million years), and lymphatic filariasis (LF) (1.616 million years).

Table 2. Global burden for 14 of 20 NTDs estimated in Disability-Adjusted Life Years (DALYs).

| Diseases | WHO Global Health Estimate (2019)[a] | | | Global Burden Disease Study (2019)[b] |
|---|---|---|---|---|
|  | YLLs | YLDs | DALYs | DALYs |
| Human African Trypanosomiasis | 101,091 | 1,009 | 102,099 | 82,600 |
| Chagas disease | 159,632 | 57,482 | 217,113 | 275,000 |
| Schistosomiasis | 428,141 | 1,199,703 | 1,627,844 | 1,640,000 |
| Leishmaniasis | 420,844 | 301,433 | 722,278 | 697,000 |
| lymphatic filariasis | 105 | 1,616,028 | 1,616,133 | 1,630,000 |
| Onchocerciasis | 12 | 1,209,707 | 1,209,720 | 1,230,000 |
| Cysticercosis | 348,191 | 639,618 | 987,809 | 1,370,000 |
| Echinococcosis | 387,710 | 73,213 | 460,923 | 122,000 |
| Dengue | 1,413,126 | 539,243 | 1,952,369 | 2,380,000 |
| Trachoma | 22 | 194,369 | 194,391 | 181,000 |
| Rabies | 2,634,634 | 146 | 2,634,780 | 782,000 |
| Soil-transmitted helminthiasis: | 170,570 | 1,772,444 | 1,943,014 | 1,970,000 |
| a) Ascariasis | 170,523 | 578,095 | 748,618 | 754,000 |
| b) Trichuriasis | 0 | 231,942 | 231,942 | 236,000 |
| c) Hookworm disease | 47 | 962,407 | 962,454 | 984,000 |
| Foodborrne trematodes | 0 | 805,406 | 805,406 | 780,000 |
| Leprosy | 7,553 | 28,884 | 36,437 | 28,800 |
| Total | 6,242,201 | 10,211,129 | 16,453,330 | 15,142,400 |

[a]Data source: **World Health Organization (2020)**; [b]Abbafati et al. (2020)

## NTDs in South East Asia (SEA)

The Southeast Asia region consists of 11 countries that are of tropical climate, in which 10 of these countries (except for Timor-Leste) make up the Association of Southeast Asian Nations (ASEAN) member states (https://asean.org/about-asean). Brunei Darussalam, Cambodia, Lao People's Democratic Republic (PDR), Malaysia, Philippines, Singapore and Vietnam are listed as Member States of WHO Western Pacific Region, whereas Indonesia, Myanmar, and Thailand are Member States of WHO South-East Region. To grasp the disproportionate burden of NTDs worldwide, we summarized the DALY for each WHO region in Table 3. Based on the DALYs estimate by cause and WHO region in 2019, dengue has the highest burden of 1.510 million DALYs, followed by LF (1.029 million years), STHs (0.616 million years), rabies (0.455 million years), and cysticercosis (0.109 million years) in WHO-SEA Region. A different trend is seen in WHO Western Pacific Region where the leading burden is rabies (0.728 million years), food-borne trematodes (0.721 million years), cysticercosis (0.376 million years), dengue (0.211 million years),and STH (0.222 million years) at the bottom of top-five NTDs DALY estimates. We discuss further NTDs of great public concern in Southeast Asia and Western Pacific region namely: Dengue, STH, Rabies, Cysticercosis, LF, and Food-borne trematodes.

Table 3. The disability-adjusted life year (DALY) Estimates for 14 of 20 NTDS by WHO Region.

| Diseases | WHO Regions | | | | | |
|---|---|---|---|---|---|---|
| | Africa | America | Eastern Mediterranean | Europe | Southeast Asia | Western Pacific |
| Human African Trypanosomiasis | 102,066 | 34 | 0 | 0 | 0 | 0 |
| Chagas disease | 0 | 216,428 | 0 | 649 | 0 | 36 |
| Schistosomiasis | 1,314,151 | 89,976 | 128,868 | 894 | 488 | 93,468 |
| Leishmaniasis | 271,033 | 36,327 | 341,843 | 3,464 | 69,590 | 21 |
| lymphatic filariasis | 412,847 | 17,432 | 22,406 | 0 | 1,028,642 | 134,805 |
| Onchocerciasis | 1,204,501 | 483 | 4,736 | 0 | 0 | 0 |
| Cysticercosis | 371,541 | 126,959 | 829 | 3,401 | 109,144 | 375,935 |
| Echinococcosis | 156,691 | 23,966 | 132,135 | 39,846 | 47,016 | 61,269 |
| Dengue | 49,712 | 76,976 | 104,969 | 24 | 1,509,709 | 210,980 |
| Trachoma | 79,158 | 1,083 | 50,443 | 22 | 50,603 | 13,082 |
| Rabies | 1,328,185 | 5,548 | 107,891 | 10,279 | 454,966 | 727,911 |
| Soil-transmitted helminthiasis: | 819,390 | 104,329 | 175,762 | 5,203 | 616,183 | 222,148 |
| a) Ascariasis | 243,250 | 35,320 | 111,253 | 930 | 303,462 | 54,403 |
| b) Trichuriasis | 57,841 | 23,285 | 4,445 | 100 | 74,341 | 71,931 |
| c) Hookworm disease | 518,299 | 45,724 | 60,064 | 4,173 | 238,380 | 95,814 |
| Foodborrne trematodes | 0 | 52,454 | 17,980 | 1,839 | 12,277 | 720,857 |
| Leprosy | 6,059 | 8,786 | 1,712 | 61 | 16,988 | 2,831 |
| Total | 6,934,724 | 865,110 | 1,265,336 | 70,885 | 4,531,789 | 2,785,491 |

Data source: World Health Organization (2020)

# Dengue

Dengue is a mosquito-borne viral disease transmitted to humans through the bites of infected female mosquitoes mainly of the species *Aedes aegypti* (Simo et al. 2019). The disease is caused by members of the genus *Flavivirus*, within the *Flaviviridae* family (Molyneux 2019). There are four distinct, but closely related serotypes of the virus that causes dengue, namely DENV-1, DENV-2, DENV-3 and DENV-4 (Braack et al. 2018). These viruses are capable of causing illnesses of dengue fever (DF), dengue haemorrhagic fever (DHF), and dengue shock syndrome (DSS) (Wibawa & Satoto 2016). Recovery from dengue infection is believed to provide immunity against that particular serotype with partial and temporary cross-immunity against other serotypes. Hence, the recovered individual is still vulnerable to dengue infection caused by other serotypes, with increased risk of developing severe dengue (Tsheten et al. 2021). Prevalence of dengue was reported in tropical and subtropical areas of Africa, America, Southeast Asia, the Pacific Ocean, and Western Mediterranean (Simo et al. 2019).

The number of dengue cases reported to WHO increased over 8-fold over the last two decades, from 505,430 cases in 2000, to over 2.4 million in 2010, and 5.2 million dengue cases reported in 2019 (World Health Organization 2021). Subsequently, the WHO Regional Technical Advisory Group for dengue and other arbovirus diseases in October 2021 reported approximately 3.5 billion people are living in dengue-endemic countries, of which 37% (1.3 billion populations) live in dengue endemic areas from 10 countries of the WHO Southeast Asia Region. Contributing factors for the widespread distribution of dengue mosquito vectors and viruses are due to high rates of population growth, inadequate water supply and poor practices, poor sewage and waste management systems, and a surge in global commerce and tourism. GBD 2019 estimated 2.38 million DALYs lost and age-standardized rates of 32.1 DALYs per 100,000 (95% UI 11.1 – 44.1) (Abbafati et al. 2020).

Currently, Dengvaxia® (CYD-TDV) is the very first licensed dengue vaccine that has been approved by the FDA and has been licensed in 20 countries. Dengvaxia® is a live attenuated, recombinant prophylactic, tetravalent viral vaccine employing the attenuated Yellow Fever virus 17D strain as the replication bone (Guirakhoo et al. 2001; Tully & Griffiths 2021). Unfortunately, the vaccine had its drawback. When issued to seronegative dengue trial recipients (individuals who have not gotten dengue infection before from any of the four serotypes), the vaccinated seronegative group exhibited similar clinical features of severe dengue to those of unvaccinated seropositive groups (people who have had dengue infection in the past) (Thomas & Yoon 2019). Additionally, the vaccinated seronegative group exhibited higher risk of plasma leakage and severe thrombocytopenia compared to unvaccinated seronegative trial participants (Sridhar et al. 2018). Due to the incidence where the vaccine induces an immune status that increases the risk of severe dengue at the time of the first natural dengue infection for vaccinated seronegative groups, WHO does not recommend the its use in seronegative individuals but endorsed it among seropositive populations based on the successful protective nature from vaccine trial studies (Tian et al. 2022). Clinical trials of the dengue vaccine were proven to be efficacious and safe when issued to seropositive recipients (World Health Organization 2018). Other than the vaccine, there is no specific antiviral treatment available for dengue illness, and the only approach to control or prevent dengue virus transmission is through interventions targeting mosquito vectors (Simo et al. 2019; World Health Organization 2022a).

## Soil-transmitted Helminths (STH)

The Ancient Greek word 'hélmins' means intestinal worm. Up-to-date, helminths are a group of parasitic worms and representative of an infection category in WHO list of NTDs comprised from the phylum *Nematoda* (roundworm) and *Platyhelminthes* (flatworm) (Akinsanya B, Adubi Taiwo, Macauley Adedamola 2021). Helminths can be further divided into three major groups of cestodes (tapeworms), nematodes (roundworms), and trematodes (flukes) based on the definitive classification of external and internal morphology of egg, larval, and adult stages (Castro 1996). Helminth infections are one of the most widespread infectious agents causing debilitating illnesses in the human population. The leading cause of human helminthiases in 2019 based on DALYs burden are caused by soil-transmitted helminths that typically infects the intestinal region, followed by schistosomiasis, LF, onchocerciasis, and food-borne trematodes (World Health Organization 2020).

STHs illnesses focused by WHO consisted of three distinct disease conditions namely ascariasis, trichuriasis, and hookworm diseases. There are four main nematode species responsible for causing these STHs infection. Roundworms (*Ascaris lumbricoides*), whipworms (*Trichuris trichiura*), and anthropophilic hookworms (*Necator americanus* and *Ancylostoma duodenale*) are responsible for infecting humans ascariasis, trichuriasis, and hookworm diseases (Mogaji et al. 2020; Zhu et al. 2020). Mode of infection in humans is through contact with parasitic eggs or larvae in soil (Brooker, Clements & Bundy 2006). Nevertheless, all STH parasitizes the intestinal tract. Ascariasis and trichuriasis are due to ingestion of fecal-contaminated food or water, or any form of fecal-oral route transmission (Muñoz-Antoli et al. 2022). Hookworm infection is transmitted primarily by walking barefoot on contaminated soil, in which the larvae mature into a form that can penetrate the skin of humans (Bethony et al. 2006).

Studies revealed that presence of high prevalence of STH are due to poor sociodemographic and socioeconomic status, especially in rural areas with poor infrastructure facilities, improper sewage and waste management, inadequate water supply, prolonged direct contact with soil such as walking barefooted, and poor sanitation and self-hygiene (Alelign, Degarege & Erko 2015; Ali et al. 2020). Anthelmintic medications (drugs that remove parasitic worms from the body), such as albendazole or mebendazole in combination with either oxantel pamoate and ivermectin, are the drugs of choice for treatment regardless of the species of STH infections (Casulli 2021). As of 2022, significant reductions of STH infections were reported through mass drug administration (MDA) as a principle of preventive chemotherapy (PC), improved water supplies and sanitation, and hygiene education programs interventions (Zeynudin et al. 2022). GBD 2019 estimated similar figures to WHO GHE 2019, with DALYs burden of 1.970 million years and age-standardized rates of 26.6 DALYs per 100,000 (95% UI 17 − 40.5). Similar trends were noted when broken into the three respective helminth diseases. Ascariasis with 754,000 DALYs burden and age-standardized rates of 10.4 DALYs per 100,000 (95% 6.6 - 15.6). Trichuriasis with 236,000 DALYs burden and age-standardized rates of 3.1 DALYs per 100,000 (95% 1.7 - 5.3). Hookworm diseases account for 984,000 DALYs burden and age-standardized rates of 13.2 DALYs per 100,000 (95% 8.4 - 19.6)(Abbafati et al. 2020).

# Rabies

Rabies is a zoonotic disease from the *Lyssavirus* genus in the *Rhabdoviridae* family that has the capability to infect all mammalian lifeforms (Condori et al. 2020). The disease has caused a high count of human mortality and economic consequences (World Organisation for Animal Health 2008). It is an acute, progressive encephalitis caused by a lyssavirus (Brown et al. 2016). It causes inflammation of active tissues of the brain which can lead to onsets of headache, stiff neck, sensitivity to light, mental confusion and seizures (Simon et al. 2013). Bite wounds or entrance of infected saliva into open cut wounds are primarily how the virus is transmitted from a rabid to healthy hosts (World Organisation for Animal Health 2008). Those bitten by dogs or inflicted with open wounds by any suspected rabid animals should seek for post-exposure prophylaxis (PEP) treatment as a necessary means to prevent human rabies infection (Bamaiyi 2015). The virus proliferates on the central nervous system (CNS) and thus laboratory techniques on sample processing are focused at said area (Sataloff, Johns & Kost 2018).

Prior to an infection, the virus enters the eclipse phase and replicates in the surrounding muscle tissues. Transmitted viruses would attach themselves to target cells through G-protein receptors and amplify in muscle tissues (Tsiang et al. 1986). Then, the virus enters peripheral nerves to be transported to the CNS. Once it has disseminated into the CNS, the virus will infect the neurons and distribute further into highly innervated tissues via the peripheral nerves. Presence of the virus can then be found in saliva and cerebrospinal fluid (CSF), nervous tissues, and salivary glands (The Center for Food Security and Public Health 2012). Two forms of rabies disease may follow which are furious rabies and paralytic rabies (Division 2018). Patients with furious rabies are easily diagnosed as they exhibit hyperactivity, excited behavior and hydrophobia, whereas patients with paralytic rabies exhibit signs of paralysis (World Health Organization 2019; World Organisation for Animal Health 2008). Nevertheless, manifestations of clinical symptoms in patients always end in death (The Center for Food Security and Public Health 2012).

An estimated 2.635 million DALYs were lost in 2019 by WHO, with the virus being prevalent in WHO African, Southeast Asia, and Western Pacific regions. Reasons for these figures could be associated with lack of control efforts, disease awareness, and appropriate measures to be taken pre- or post-exposure prophylaxis (Schlottau et al. 2017). In contrast to the global burden DALYs estimate for rabies in 2019 by WHO, GBD 2019 estimated a lower DALYs burden of 782,000 years with age-standardized rates of 10.6 DALYs per 100,000 (95% UI 4.4 – 14.7) (Abbafati et al. 2020).

## Cysticercosis

Cestode infection involving *Taenia solium* results in two distinct illnesses. Taeniasis is an intestinal adult tapeworm infection which occurs when one consumes raw or undercooked contaminated pork. Tapeworm eggs are spread when the carrier defecates in open areas (WHO 2014). Additionally, gravid proglottids, a mature segment of the tapeworm that may detach and migrate to the anus and pass as feces, contain *T. solium* eggs and is considered as another means of spreading the infection (Jansen et al. 2021). Human taeniasis is often asymptomatic, but mild symptoms may manifest which includes abdominal pain, distention, diarrhea, and nausea (García et al. 2003).

Cysticercosis is an infection in both humans and porcine, caused by parasitic larval form (cysticercus), after consuming food or water contaminated with feces containing *T. solium* eggs (fecal-oral contamination) (Lustigman et al. 2012). Once ingested, the eggs hatch in the intestine, releasing oncospheres that invade the intestinal wall and entering the bloodstream, and then migrating to multiple tissues and organs (muscles, skin, eyes, and central nervous system) where they then mature into cysticerci (Galipó et al. 2021). Development of parasitic cysts in the brain or central nervous system is referred to as neurocysticercosis (NCC) (Laranjo-González et al. 2017).

Both diseases are predominantly prevalent in endemic countries with inadequate hygiene practice and poor basic sanitation measures. As cycticercosis is acquired by ingesting *T. solium* eggs shed in the feces of a human *T. solium* carrier, this illness can still occur in populations that neither eat pork nor share environments with pigs, as long as there is a human carrier present. WHO reports 51 countries with endemicity of *T. solium* in 2022 compared to the 2018 report of 57 endemic countries (https://www.who.int/data/gho/data/themes/topics/taeniasis-cysticercosis). In addition to the global burden DALYs estimate for cysticercosis in 2019 by WHO, GBD 2019 estimated a higher DALYs burden of 1.370 million years with age-standardized rates of 16.8 DALYs per 100,000 (95% UI 10.7 – 23.9) (Abbafati et al. 2020).

## Lymphatic filariasis (LF)

LF is caused by a group of helminths (roundworm) from the family of *Filariodidea* that reside in the lymphatic systems of humans (Wibawa & Satoto 2016). Majority of the infections worldwide are caused by *Wuchereria bancrofti*, followed by *Brugia malayi* and *Brugia timori*. (Mitra & Mawson 2017). A wide range of mosquito species are responsible for the spread of LF and can be concluded in the primary genera of *Anopheles*, *Culex*, *Aedes*, and *Mansonia* (Famakinde 2018). Similar to many mosquito-borne diseases, LF infection is transmitted when mosquitoes pick up the microfilariae during a blood meal of an infected LF individual, the parasite develops within the mosquito and subsequently infects the next victim on the subsequent blood meal (Douglass et al. 2017).

Disease morbidity results from damage to one's lymphatic vessels by adult parasite nests and microfilaria released in the bloodstream. Functionally impaired lymphatic systems will lead to manifestation of lymphoedema (elephantiasis), hence the enlarged state of the patient's limbs. Physically impaired people experience years of disability, stigmatization, and mental health comorbidity (Abela-Ridder et al. 2020). LF is one of the significant causes of permanent disability worldwide with an estimated 1.616 million DALYs lost in 2019, and still affecting more than 1 billion people globally in 2021 (population requiring MDA and treatment for LF) (World Health Organization 2022). When LF develops into chronic conditions it leads to lymphoedema (tissue swelling) or elephantiasis (skin/tissue thickening) of limbs and hydrocele (scrotal swelling) (Lupenza, Gasarasi & Minzi 2022). The Global Programme to Eliminate Lymphatic Filariasis (GPELF) was launched in 2000 by WHO in response to the call by World Health Assembly resolution WHA50.29 (World Health Assembly 1997; Yajima & Ichimori 2021). GPELF is motivated with the primary goal to interrupt LF transmission through annual mass drug administration (MDA) with albendazole in combination with either ivermectin in Africa or diethylcarbama-zine in areas outside Africa, and the secondary goal of alleviating the morbidity burden associated with LF infection (Dickson, Graves & McBride 2017). GPELF has achieved great successes since its inception, with the prevalence reduction of 74% between 2000 to 2018, along with validated status elimination of LF in 17 countries (Casulli 2021; Hotez & Lo 2020). GBD 2019 estimated a slightly higher DALYs burden of 1.630 million years compared to WHO GHE 2019, with age-standardized rates of 20.7 DALYs per 100,000 (95% UI 12.2 – 34.7) (Abbafati et al. 2020).

# Emerging neglected diseases.

## Melioidosis

Melioidosis or Whitmore's disease is a bacterial infection caused by *Burkholderia pseudomallei*, a soil- and water-borne Gram-negative bacterium capable of causing illness ranging from an acute or chronic localized infection to a widespread septicemic infection in multiple organs (Cheng & Currie 2005). Cases of melioidosis are frequently reported in endemic countries such as Africa, Australia, China India, Middle East, and Southeast Asia (typically Malaysia, Singapore, and Thailand) (Cheng & Currie 2005; Galyov, Brett & Deshazer 2010). Since its discovery in 1912, this bacteria still remains a topic of discussion among researchers due to being zoonotic in nature, limited therapeutic options with no available vaccines till date, making the etiological agent capable of causing economic crises at unpredicted outbreaks (Borlee et al. 2017). The United States Select Agent designated *B. pseudomallei* as a Tier 1 agent due to their biothreat potential including high morbidity and mortality rates in low infectious doses, multidrug antibiotic resistance, and the amenability to be aerosolized (Hatcher, Muruato & Torres 2015).

Melioidosis infection can be acquired through many routes with skin inoculation and inhalation or ingestion of contaminated water and air droplets to be the leading cause (Larsen & Johnson 2009). Disease severity varies depending on the bacterial mode of infection, strain, as well as the host's susceptibility and immunological state (Chakravorty & Heath 2019; Wiersinga et al. 2018; Wolff et al. 2021). Among many melioidosis physiology, pulmonary melioidosis infection confers greater lethality compared to non-pneumonic type and lung involvement were observed in approximately 50% of the patients particularly during the rainy seasons due to direct inhalation of the bacterial particle. Inhalational infection causes the most severe damage to the host due to the rapid dissemination rate to other vital organs such as spleen and brain (Cheng et al. 2008; Limmathurotsakul et al. 2005). This imposed a great occupational risk among farmers as they are prone to minor cuts to their feet and hands along with prolonged exposure to muddy water in the paddy field (Chaowagul et al. 1989). Prior to first clinical manifestation of melioidosis, the bacteria may lay dormant ranging from days to years to evade the detection of the host immune systems awaiting the opportunity for relapse (Pal et al. 2022). Furthermore, melioidosis mimics the signs and symptoms of other diseases (tuberculosis, malaria, dengue) often complicating the accurate diagnosis for the disease (Sarkar-Tyson & Titball 2009). As such, a study on the burden of melioidosis has estimated approximately 165,000 cases with a >50% mortality rate (89,000 deaths) globally in 2015 (Limmathurotsakul et al. 2016). There is no available data reporting DALYs estimate for melioidosis from WHO GHE 2019 and GBD 2019. Up-to-date presently available global burden of melioidosis in 2015 by Birnie et al. (2019) described an estimated 4.64 million DALYs which surpassed all 16 other NTDs.

## Leptospirosis

Leptospirosis is a zoonotic disease caused by a lethal bacteria of the genus *Leptospira* (Horwood et al. 2019). In the host, the bacteria reside in the kidney to undergo its lifecycle and are then shed in the urine. A molecular serotyping study concluded a more than 20 *Leptospira spp.* which can be further segregated into three phylogeny clades of pathogenic, intermediate, and non-pathogenic (Levett 2015).

Various wild and domesticated mammals can act as host reservoirs for *Leptospira spp.* in the city; rodents are considered as one of the most important host source of leptospirosis infection as they can persistently shed pathogenic *Leptospira spp.* to the environment throughout their lifecycle without any clinical manifestations (Urbanskas, Karvelienė & Radzijevskaja 2022). Rapid urbanization and population growth in the city along with poor sanitation and waste management by the city council propels the prevalence and spread of *Leptospira spp.* by city rats (Dobigny et al. 2018). Plantations are a rich source of food for rodents which further favours the presence of rats thus increasing the chance for the transmission of Leptospira spp. through urinal discharge at places with high human mobility (Garba et al. 2018). The *Leptospira spp.* has great adaptability in both environmental and within the host reservoir. Andre-Fontaine, Aviat & Thorin (2015) reported Leptospira survival of as long as 10 months in 4°C and up to 20 months when kept at 30°C. Formation of biofilm by the bacteria aids in evading the hosts' immune system detection, reducing signaling activity to the immune system resulting in low titres of antibodies in the hosts (Almeida et al. 2019).

Human individuals can contract the illness through direct contact with Leptospira-contaminated urine, water, and wet soil (Sun, Liu & Yan 2020). Individuals infected with pathogenic *Leptospira spp.* may be asymptomatic or associated with varying clinical manifestations ranging from acute febrile illness to severe characterized dysfunction of multiple organs leading to death (Sykes et al. 2022). Patients would exhibit sudden onset of fever, chills, and headache which mimics the symptoms of other threatening diseases such as dengue, influenza, and malaria (Haake & Levett 2015). Patients may recover from the symptoms but if left undetected, a second more severe phase will occur leading to kidney or liver failure, and possibly meningitis (Abdullah et al. 2019). In countries or regions where diagnosis confirmation by laboratory tests is limited, this causes the number of reported leptospirosis to be underreported and thus neglected. Due to this, there is no accurate data available from WHO GHE and GBD. To gain a grasp of the global burden of leptospirosis, a model study estimated approximately 1.03 million cases of leptospirosis worldwide annually, of which 5.72% (58,900) results in death (Costa et al. 2015). Additionally, those figures were incorporated by Torgerson et al. (2015) to estimate the global burden of leptospirosis in terms of DALYs which were predicted to be at 2.90 million DALYs annually, representing incidence of 41.8 DALYs per 100,000 population (UI 18.1 - 65.5).

## Malaria

Malaria is an ancient life-threatening disease caused by parasites transmitted through the bites of infected female *Anopheles* mosquitoes (Christophers 1951). The causative agent for malaria is a group of unicellular protozoan parasites originating from the *Plasmodium* genus (Sato 2021). All *Plasmodium* spp. are capable of infecting malaria but to a specific range of host, and there are *P. falciparum*, *P. vivax*, *P. malariae*, *P. ovale*, and *P. knowlesi* that are naturally capable of infecting humans with malaria (Lalremruata et al. 2017). In addition, two of the aforementioned *Plasmodium* spp. are of great research focus as *P. falciparum* is the deadliest and most prevalent malaria parasite on the Africa continent whereas *P. vivax* is the most dominant malaria infection outside of sub-Saharan countries (Larson 2019; Liu et al. 2014).

There were an estimated 241 million cases of malaria in 2020, and the estimated number of malaria deaths stood at 627,000 (Singh et al. 2022). In the same report, nearly half of the world's population was at risk of malaria with the most cases and deaths reported in sub-Saharan Africa. However, the WHO regions of South-East Asia, Eastern Mediterranean, Western Pacific, and the Americas also report significant numbers of cases and deaths. The WHO African Region carries a disproportionately high share of the global malaria burden (https://www.who.int/data/gho/data/indicators/indicator-details/GHO/estimated-number-of-malaria-cases). In 2020, the region was home to 95% and 96% of malaria cases and deaths, respectively. Children under 5 years of age are the most vulnerable group affected by malaria; in 2020, they accounted for about 80% of all malaria deaths in the WHO African Region (Saba, Balwan & Mushtaq 2022). GBD 2019 estimated a significantly higher DALYs burden of 46.4 million years compared to WHO GHE 2019, with age-standardized rates of 667 DALYs per 100,000 (95% UI 337 − 1,150) (Abbafati et al. 2020). .

## Surveillance and Disease Management

The DALYs burden estimates for each of the aforementioned diseases underlines a pressing need for a clear guide and protocol by government authorities and international bodies in reaching elimination targets for these diseases. Here, we review surveillance programmes and disease management actions that are being implemented in response to these neglected diseases.

Dengue surveillance is crucial for detecting outbreaks and monitoring disease incidences. Increasing the number of surveillance traps that capture eggs (ovitraps) and ovipositing females (gravid traps) with appropriate larvicide and mosquitocide (Selvarajoo et al. 2022). This is to prevent hatching of eggs or any subsequent production of mosquitoes inside the trap. This method is a double prong approach allowing authorities to survey the incidences and population of mosquitoes as well as for vector control. However, counting of both traps requires a group of individuals (insectaries) that possess considerable degree of skill to count eggs and sampling specific stages of mosquitoes (Tsheten et al. 2021). Other than vector surveillance, governing authorities should also carry out susceptibility bioassays to detect presence of insecticide resistance in the vector population. Observation of the phenotypic response of mosquitoes post-exposure to insecticides should be a sufficient metric in determining presence of insecticide resistance.

Control of dengue is mainly achieved through the cooperation of all walks of life. Common strategies include removal of mosquitoes breeding sources, eliminating container habitats that would collect water that are favorable for oviposition sites and development of mosquito larvae, killing of larval or pupal mosquitoes by applying environmentally friendly insecticides, and usage of spatial repellents (Srisawat et al. 2022). Various strategies of sterile insect technique (SIT) all aimed at causing decline of targeted insect population through the release of sterilized male insects were proven to be a significant measure in controlling mosquito populations (De Castro Poncio et al. 2021; Hugo et al. 2022; Ranathunge et al. 2022). Despite these advances, dengue infection remains largely uncontrollable in both rich and poor populations due to several factors. This was seen in Singapore's 15 year-long intensive vector surveillance from the mid-1970s to the late 1980s, where low incidences of dengue were reported. In 1990 onwards, the country faced repeating cyclical epidemics where the largest epidemic occurred back in 2013 with over 22,000 cases despite the continued investment of US$50 million in vector control annually (Molyneux 2019). Hence, there is a need for improving active surveillance aspects and developing drugs for better disease management.

Since no vectors are involved in the transmission of STHs illnesses among humans, surveillance programmes can be separated into (i) prediction- and mathematical-based models covering transmission model, estimating of population at risk, predicting regions in need of MDA interventions, and (ii) active surveillance approach (Chong et al. 2022; Mogaji et al. 2022). Soil and stool samples are important determinants of the susceptibility level of a population to STHs (Oyewole & Simon-Oke 2022). The former includes usage of algorithms and large data inputs to make precise decision making of incidences and prevalence for STHs in endemic areas. In contrast, the latter involves the expert skills of laboratory staff and technicians to conduct microscopy-based and molecular-based detection methods to detect presence of helminths and accounts for disease surveillance at the same time . Microscopy-based methods, such as the Kato-Katz and McMaster techniques, are primary diagnostic tests to detect parasites by enumerating the eggs-per-gram (EPG) metric. Although these methods are cheaper, results may vary with an increase of sample size and different survey sites (Afolabi et al. 2022). Hence, molecular-based assays such used for diagnosis such as PCR, real-time PCR and digital PCR, loop-mediated isothermal amplification (LAMP), and cell-free DNA detection provided a more sensitive, less labor-intensive, and high-throughput detection method despite incurring additional costs for detection and surveillance programmes (Becker et al. 2018; Manuel, Ramanujam & Ajjampur 2021).

Measures taken in managing cases of STHs are shared among other helminth infections. Implementing MDA programs targeting high-risk groups in endemic tropical and subtropical areas has been recognized to be effective in eliminating STH globally (Alemu et al. 2022). In underdeveloped areas with poor facilities but endemic with STHs, adopting the principles of WASH by providing adequate sanitation, improving waste management facilities, plus public education on hygiene practices and behavioral changes targeted to populations at risk would accelerate in achieving the elimination goal for good (Sato et al. 2019). As for those diagnosed with STHs, proper provision of treatment (stronger combination of antibiotics or surgery to remove the worms) is essential in ensuring a good universal health coverage (Abela-Ridder et al. 2020).

Rabies surveillance and monitoring are largely conducted via timely detecting of animals showing suspected clinical signs of rabies, recording the history of recently dead companion animals, and follow-up on dog-bite cases (Dadang 2019). In response to the 2017 rabies outbreak that occurred in Malaysia, the authors have listed in detail rabies preventive measures, and control procedures for outbreak (Dadang 2019). Generally, the best method to control cases of rabies is to visit the nearest healthcare or veterinary services to get pre-exposure prophylaxis for both you and your companion animals. Rabies surveillance-diagnosis programmes are challenging as the gold standard for rabies lyssavirus detection is direct diagnosis with brain tissue. Next, efforts in controlling the disease are troublesome when incidences of free roaming stray dogs and cats are considered a norm, plus close vicinity to wildlife habitat. Hence, most countries would only initiate a mass dog vaccination programme to control the outbreak and to curb the transmission (Molyneux 2019).

Surveillance programmes for rabies were proven to be challenging as rabies-infected (rabid) animals can only be identified when clinical symptoms manifest and often lead to death. Hence, drastic actions are needed to reduce rabies mortality. As children are more vulnerable to animals, education is crucial in preventing deaths from rabies. Lessons on how to avoid being bitten and what to do in the event of a suspected bite from a rabid animal reduces mortality rate among children remarkably. Immediate washing of the bite wound can significantly decrease the viral load in the wound. To break the cycle of rabies transmission, implementing mass vaccination for dogs by health authorities has been recognized to be more cost-effective and protects the well-being of livestock and humans at the same time. As mentioned before, vaccination (pre- or post-exposure prophylaxis) is highly recommended for people at high risk of exposure for rabies such as veterinarians and laboratory staff working closely with rabies virus, and those bitten by a potentially rabid animal (Abela-Ridder et al. 2020).

A virtual meeting convened by WHO aimed at reviewing existing diagnostic tools for *T. solium* before implementing them as public health programmes to control the disease (World Health Organization 2022b). In the context of a programmatic surveillance, WHO recommended inclusion of specific communities or villages that are of a wider geographical area. Purposive sampling should be implemented to target high-risk humans and pigs' interactions when the community lives in close contact with pigs or pigs roam freely where sanitation is inadequate. All means of diagnostic mapping and monitoring of *T. solium* presence in humans and pigs did not achieve the required sensitivity due to confirmatory methods via microscopy (humans), meat inspection, and serology testing. Diagnostic tests for public health programmes are not well-suited as they are not commercially available and of unsatisfactory sensitivity and/or specificity. Due to the low sensitivity of currently available tests, WHO urged for an appropriate response even if it's a weak signal of prevalence detected. Preventive chemotherapy (PC) interventions should be initiated with confirmation of key risk factors of roaming pigs and inadequate sanitation.

Cysticercosis can be successfully eliminated as a public health concern as interventions in disease management are feasible and achievable. Beginning with *T. solium* in pigs, improving the well-being quality and management of pig husbandry can effectively break the transmission cycle. Such actions are such as vaccination programs, anthelmintic treatment for pigs, and proper set up of enclosure habitats to prevent access to human feces. To curb incidences of taeniasis and cysticercosis among humans, practice of proper WASH principles such as proper sanitation for safe disposal of feces, and proper hygiene and food safety would significantly reduce chances of contracting the diseases. In addition, community health education, MDA interventions, and appropriate case management for taeniasis (medical prescriptions) and NCC (surgery) would aid in removing the disease status as a public health concern (Abela-Ridder et al. 2020).

Establishment of GPELF by WHO had set the goal in managing the transmission of LF infections via MDA of anthelmintics and alleviating the sufferings of people affected through morbidity management and disability prevention (MMDP). Surveillance programs for LF were conducted through sentinel and spot-check community surveys. Periodic transmission assessment survey (TAS) measures the impact of MDA interventions and to determine if the level of infection decreases below a target threshold (World Health Organization 2022c).

Approaches in LF disease management include MDA intervention to stop the spread of infection through WHO-recommended prescriptions of albendazole, diethylcarbamazine, and ivermectin. Usage of insecticide-treated bednets is recommended as a means of vector control in household environments. Integration of WASH could be applied to ensure hygiene care of affected limbs for morbidity management, and proper sanitation practices to reduce vector breeding sites. As LF is known to cause significant physical impairment, health authorities must ensure essential care is given to patients. Examples are such as skin care, exercises and elevation to prevent severity and progression of lymphoedema, treatment for episodes of adenolymphangitis, surgery to cure hydrocele, and promoting collaboration among communities to complete treatment and cope with physical and mental consequences (Abela-Ridder et al. 2020).

In summary, nationally representative survey programs suited for the geographical and environmental etiological factors for each respective country, such as demographic and health surveys (DHS), may offer an appropriate platform for active disease surveillance. Picking the proverb "prevention is better than cure", core strategic interventions together with disease management will better facilitate in eliminating the prevalence and transmission of diseases, and at the same time decrease the morbidity and mortality inflicted.

## Application of Machine Learning Tools for NTDs.

The conventional approach to drug discovery costs a fortune and takes up a considerable amount of time. Computational approaches to drug discovery using Artificial Intelligence (AI) can resolve both concerns and speed up the process of novel drug discovery. Machine learning (ML) is a subfield of Artificial Intelligence (AI) where sets of data and algorithms (mathematical and statistical) are utilized in search of distinct patterns within the data for a more efficiently accurate downstream analysis (McComb, Bies & Ramanathan 2022). ML in drug discovery and development are carried out by looking for patterns in sets of molecules with drug- and therapeutic-properties to describe in detail their biological activities (Dara et al. 2022). ML techniques can be tasked for prediction of continuous values (regression), prediction of classes (classification), and grouping of similar data items (clustering) (Oguike et al. 2022). In this section, we explore the applications of ML tools in developing drugs for a selection of NTDs such as dengue, malaria, and leptospirosis. Next, we discuss how advances in adjacent fields of protein- and antibody-language models, cancer research and computer vision can be leveraged for NTDs research and disease management. Lastly, we discuss steps taken for regional collaboration, data and infrastructure sharing within and around SEA and Western Pacific regions.

### Dengue

Khalid and coworkers (2020) reported a study that uses ML methods to investigate the biological activities of inhibitor derivatives anti-dengue compounds. They employed an atom-based three-dimensional (3D) QSAR modeling study along with the machine learning software Schrödinger Drug Discovery Suite Phase™ to investigate the compound's structural features with the anti-dengue activities. Homologous series of 21 novel 1,3,4-oxadiazole derivatives were used as training dataset. Prepared datasets were then divided into both training and test sets with relative percentages of 75% and 25% using the random selection tool integrated in Schrödinger Phase™ software. The model's predictive power based on predictability of biological activities of test molecules Q2 (R2 Training Set) = 0.73 and Q2 (R2 Test Set) = 0.78 displayed satisfactory predictive power. They successfully developed and validated their ML model that can reveal correlation of the molecule's structural relationship with the it's bioactivity (Khalid, Rao Avupati & Hussain 2020).

Dengue Virus (DENV) NS2b-NS3 protease complex is essential for the viral replication process making it a great target for antiviral agents. However, available choices of inhibitors during that time were unsatisfactory due to weak activity or low selective index towards the NS3 active site. The NS3 protease domain is essential in processing the DENV polyprotein for the replication process. The presence cofactor NS2b is significant for substrate recognition and in maintaining the complex stability of NS2b-NS3 assembling complex for the DENV replication to take place. Aguilera-Pesantes et al. used ML methods to identify potential residues and sites for drug-like molecule interaction, and bindable sites for drug development. They used four ML models, Random Forest (RF), Least absolute deviation tree (LAD Tree), voting feature interval (VTI), and multilayer perceptron algorithm (MLP), to classify their data. They found that MLP models work best in their study to properly classify residues interacting with NS3 that would cause major change in activity, moderate change in activity, and residues with similar activity as wild type residues (Aguilera-Pesantes et al. 2017).

Geoffrey et al. (2020) implemented ML-based AutoQSAR that involves feature selection, QSAR modeling, validation, and prediction to generate drug leads from PubChem database for both Dengue and West Nile virus drug studies. The program aids in speeding up virtual drug screening and identification against Dengue and West Nile virus, generating drug lead through their ML-based AutoQSAR algorithm, and automates *in silico* analysis of the drug lead compounds.

The application of Artificial Neural Network (ANN) to predict Dengue-Human protein interaction type leading to development of antiviral drugs was reported by Jainul Fathima et al. (2019). They used the Feed Forward Back Propagation Neural Network (FFBPNN) technique to train the model with a dataset consisting of 535 non-redundant interactions between 335 different human proteins and ten dengue proteins that are composed of eight attributes and 550 instances. The constructed neural network based on the training data was tested with the testing data to predict the accuracy of the BPNN model. The model prediction reported 98.05% accuracy for the dataset study.

Elakkiya Elumalai (2022) reported a study that used ML methods to classify dengue virus-inhibiting and non-inhibiting peptides. A dataset of 100 experimentally validated dengue virus inhibiting peptides and 16 negative datasets from antiviral peptides database (AVPdb), were both split into a 7:3 ratio of training and testing set. ML algorithm models employed in the study consists of. Three amino acid descriptors namely Amino acid composition (AAC), grouped amino acid composition, transition, and distribution (GAAC), and Composition, transition and distribution features (CTDC) were compared using eight different ML algorithm models Random forest (RF), Multi-layer perceptron (MLP), Support Vector Machine (SVM), Logistic regression (LR), k-Nearest Neighbour (k-NN), Naïve Bayes (NB), Adaptive Boosting (Adaboost) and Bagging. Five of their best models on training data reported accuracies greater than 85%. The same five models were used for testing in which two models (AAC_k-NN_model and AAC_RF_model) reported accuracy of 85.71%, whereas the remaining models are less than 80% accuracy. Both k-NN and RF algorithms implemented were validated as the best algorithm in achieving the research goal. In addition, their study discovered higher frequency of glycine (G), phenylalanine (F), and tryptophan (W) amino acids found in dengue virus inhibitory peptides. (Elumalai 2022)

## Malaria

Interpretable scoring card method was used to identify the antimalarial activity by using only the peptides sequence information (Charoenkwan et al. 2022). The authors trained an iAMAP with SCM-based predictor with eight other conventional supervised classifiers of decision tree (DT), k-nearest neighbor (KNN), logistic regression (LR), multilayer perceptron (MLP), naïve Bayes (NB), partial least squares regression (PLS), random forest (RF), and support vector machine (SVM) with nine conventional feature descriptors namely amino acid index (AAindex), PCP, amino acid composition (AAC), composition, transition and distribution (CTD), CTD-composition (CTDC), CTD-distribution (CTDD), CTD-trancomposition (CTDT), dipeptide composition (DPC), and tripeptide composition (TPC). The training data set consists of 139 positive and 2,135 negative molecules and the test set of 139 positive and 677 negative molecules. The program first estimates the propensities of 20 amino acids and 400 dipeptides which were then optimized via the 10-fold cross-validation scheme. Estimated propensities were then utilized to select important physicochemical properties whereas the optimal propensities of the 400 dipeptides were proceeded to develop the final predictive model (iAMAP-SCM). Scoring based on maximum accuracy (ACC) and Matthew's coefficient correlation (MCC), iAMAP-SCM was reported to achieve scores of 0.957 and 0.834 respectively and outperformed the other three classifiers employed, when the model was screened independent test datasets for validation.

Mswahili et al. (2021) developed and compared the performance of five ML models to predict antimalarial bioactivities against *P. falciparum*. They trained ML models of artificial neural network (ANN), SVM, RF, extreme gradient boost (XGB), and LR over a data set of 4,794 antimalarial drug candidate compounds (2,070 active and 2724 inactive molecules). The K-best filter-based algorithm that selects potential features according to a particular function and Recursive Feature Elimination (RFE) wrapper-based algorithm that treats feature selection as a search problem were chosen as feature selection algorithms for performance examination and comparison. K-best was adopted as an accuracy metric whereas RFE was viewed as an efficiency metric. Based on the two metrics, they found that XGB, ANN, and RF models gave the best three accuracies in finding new antimalarial drug formation without losing too much precision.

Four ML classification algorithms, namely NB, NN, RF, and SVM were employed to investigate protein-protein interaction (PPI) networks for human and malarial parasites obtained from STRING database (version 11.0) was reported by Apichat Suratanee and colleagues to identify new human proteins associated with malaria as a means for additional drugs development (Suratanee, Buaboocha & Plaimas 2021). The ML models were trained with a data set of 12,038 human proteins with 313,359 interactions, and 1,787 *P. vivax* proteins with 11,477 interactions. They constructed a heterogeneous network connecting human-human protein interactions and *P. vivax-P. vivax* protein interactions with the human-*P.vivax* protein associations while investigating five topological features of (i) betweenness centrality, (ii) closeness centrality, (iii) degree, (iv) eccentricity, and (v) Kelinberg's hub centrality. They applied ten 10-fold cross-validations for each algorithm to yield performance metrics of an ROC curve with an AUC in which RF algorithm was the best classifier (AUC of 0.85) followed by NN (AUC of 0.79), SVM (AUC of 0.77), and NB (AUC of 0.74). With the best performance and results of the RF classifier, the authors obtained 411 human proteins through a top-ranking score calculation for each human protein. Subsequent functional annotation of the proteins revealed previously reports of promising candidates for multistage targets for malaria therapy.

## Leptospirosis

Abdullah et al. (2021) studied the identification of a suitable *Leptospira spp.* multiepitope-based vaccine candidate which utilized two ML programs, namely Vaxign-ML and C-ImmSim. In his study, all protein antigens have protegenicity score greater than 90% signifying as effective antigens for vaccine developments, and simulations from C-ImmSim showed diverse immune reactions of the vaccine construct indicating promising subunits of multiepitope vaccine candidate for immunity against *Leptospira spp.* Infections.

Vaxign-ML is a supervised ML classification reverse vaccinology (RV) program trained to predict rank score (termed protegencity) of bacterial protective antigens (BPAgs) based on a training data set consisted of viral and bacterial antigens (Ong et al. 2020). Through a nested 5-fold cross-validation (N5CV) and leave-one-pathogen-out validation (LOPOV) evaluation approach, extreme gradient boosting (XGB) was the best out of five other ML algorithms. Set as the benchmark against five other existing programs and methods, Vaxign-EGB-ML displayed satisfactory results outperforming four programs. Final validation on external data sets of clinical trials or licensed vaccines reported ranked calculation of best top 10% BPAg candidates for 20 proteins. Next, C-ImmSim is an immune-simulation study server that employs machine learning methods and position-specific scoring matrices to identify epitope peptides and other immune interactions (Rapin et al. 2010). The program combines a mesoscopic scale simulator of the immune system with a set of agent-based class computational models to predict molecular-levels of major histocompatibility complex-peptide binding interactions and neural networks for prediction of epitopes.

## Comparisons with ML application in cancer research, computer vision, protein language models

Majority of studies employing ML models to discover novel drug candidates for NTDs have been published in the past two decades. Similarly, applications of ML in cancer research have been in practice since the early 2000s (Bertsimas & Wiberg 2020). Research domains where ML-based methods can be employed in cancer biology includes genomics, proteomics, metabolomics, epigenetics, transcriptomics, and system biology (Kourou et al. 2021; You et al. 2022). ML tools developed specifically for the molecular study of cancer have. An overview of the application of AI in identifying cancer targets and drug discovery has been reviewed (Alqahtani 2022; Shao et al. 2022; Taylor 2020; You et al. 2022). Despite the advances in chemotherapy and immunotherapy, early detection of cancer increases one's survival rate tremendously. Compared to detection at later stages, the cancer would have metastasized, spread to vital organs where surgery may not be feasible and hence has been sentenced to doom. Thanks to technological innovations, a new branch of AI known as computer vision (CV) will significantly lighten the burden of physicians and radiologists when it comes to interpreting an MRI or histology slide for the presence of a tumor.

In the past six years, an increasing array of ML tools has been developed in response to cancer diagnosis as well. Literature mining on ML-based studies on cancer diagnosis, patients' classification, and prognosis (excluding reviews and technical reports) between 2016 and 2020 in PubMed biomedical repository and Digital Bibliography and Library Project (DBLP) computer science bibliography yielded 921 and 165 studies respectively. Additionally, the total number of articles started with around 25 in 2016 and ending with approximately 625 studies published in 2020 (Kourou et al. 2021). Advances of AI and ML techniques, particularly Deep Learning (DL) which is a subset of ML technique, can be developed to mimic human-like capabilities for data processing to identify images, objects, process languages, improve drug discovery, upgrade precision medicines, improve diagnosis, and assists in decision making with or without human supervision (Davenport & Kalakota 2019). The multi-layered neural network architecture of DL enables models to grow at an alarming rate, provided with abundantly dimensional data (Lecun, Bengio & Hinton 2015). As reviewed by Kourou et al. (2021), most cancer ML-based studies on cancer detection and diagnosis centered around imaging data (input) from computed tomography (CT), magnetic resonance imaging (MRI), X-ray radiography, and positron-emission tomography (PET) to develop DL architectures of automated diagnostic models. Successfully early diagnosis of breast cancer using convolutional neural network (CNN) to analyze histopathological images were reported, with additional validation from other researchers of promising plus accurate diagnostic capabilities of deep CNN architectures by analyzing imaging slides. Efforts to develop an image-based lung cancer detection model, a region-based CNN model trained with 42,290 whole-CT lung scans has outperformed the average radiologists at malignancy risk-prediction, and achieved AUC score greater than 95% when validated with 1,139 clinical cases (Ardila et al. 2019).

Advances in the field of AI, typically ML and DL methods, were utilized to develop language models to predict proteins. Algorithms from these methods were employed to process the efficiency and quality of the natural language processing (NLP). To develop a protein language model (PLM), large text (protein sequences from large databases) are given as input to train the prediction of masked or missing amino acids (Bepler & Berger 2021; Ofer, Brandes & Linial 2021; Rives et al. 2021). At the end of protein information processing, involving multi-dimensional vectors and hidden layers, a representative group of proteins are acquired and are referred to as embeddings as mentioned by Elnaggar et al. (2021). Advances of embeddings from literature findings displayed stunning performances in predicting secondary structure and subcellular location comparable to other methods that employ evolutionary information from MSA inputs, substituting sequence similarity for homology-based annotation transfer, and predicting mutational effects on protein-protein interactions (PPI) (Alley et al. 2019; Heinzinger et al. 2019; Littmann et al. 2021; Stärk et al. 2021; Zhou et al. 2020). Variant Effect Score Prediction without Alignments (VESPA) is able to predict sequence residue conservation and single amino acid variants (SAV) almost as comparatively accurately to other existing methods (ESM-1v, DeepSequence, and GEMME) without employing multiple sequence alignment (MSA) approach (Marquet et al. 2022).

PLM has extended its range to analyzing protein via MSA data approach as well. Homologous proteins indicate descendant relationship with an ancestral protein and thus share similar structure and function. Analyzing MSAs data of said homologous proteins would provide valuable information about functional, structural, sequence conservation and evolutionary information that the organism or gene underwent. Successful breakthrough in atomic-resolution structure prediction problem by PLM-based structure prediction models, such as AlphaFold2 (AF2) and RoseTTAFold, was achieved on the use of MSAs and templates of similar protein structures to achieve the best optimal structural prediction performance (Baek et al. 2021; Jumper et al. 2021).

Lupo, Sgarbossa and Bitbol (2022) investigated the performances of MSA-based LMs in isolating coevolutionary signals encoding functional and structural constraints from phylogenetic correlations through a set of pre-trained synthetic MSAs generated from Potts models. Three attention based neural network architecture MSA Transformers were studied, namely AlphaFold2 (AF2), RoseTTAFold and RGN2. All programs displayed efficient performance in differentiating correlations from contacts and phylogeny, which are always both present in natural data causing phylogeny noise and proven to be a fundamentally hard problem. The results demonstrated that contact inference by MSA were less deteriorated by phylogenetic correlations and showed greater accuracy in structural contacts compared to Potts models despite the MSA Transformer being pre-trained with a dataset of minimized diversity.

The substantial performance of AF2 in the recent CASP14 prediction challenge depicted the combination of MSAs data and ML-based methods is able predict unsolved protein model structure with remarkable accuracy complementing results from experimental works (X-ray crystallography, cryoEM, and NMR) (Jumper et al. 2021). It is important to note that this does not conclude protein structure prediction models implementing ML-based techniques can entirely substitute the experimental methods, but instead assists in understanding the folding nature of protein biophysics. For these programs to be at its peak (achieve high accuracy), an enormous yet varying amount of MSAs database is required to be fed in order for each of their related ML-based algorithm systems to learn and predict protein structure based on the co-evolutionary relationships encoded within the MSAs (David et al. 2022). Hence, when tasked  with predicting proteins that lack available data on sequence homology, these programs may be less performative (Pearson 2013). Thus, Chowdhury et al. (2021) has developed an end-to-end differentiable recurrent geometric network (RGN2) that is capable of predicting structure from single protein sequences without using MSAs data. The program employs two novel elements of (i) AminoBERT PLM that uses Transformer units to learn latent structural information from millions of unaligned proteins, and (ii) geometric modules representing the Cα backbone geometry. Chowdhurry and colleagues described the performance of RGN2 surpassing AF2 and RoseTTAFold when tasked with predicting proteins with no known homologs and can compete on *de novo* designed proteins. Absence of MSAs-element in RGN2 allowed protein structure prediction speed of up to six-fold faster than programs that require it.

A subsequent study on the advances of PLM in ML-based protein structure prediction since AF2 and RoseTTAFold were conducted by (Lin et al. 2022). In their study, they developed ESMFold that can compete with AF2 and RoseTTAFold in atomic level protein structure prediction accuracy with information on individual sequence of a protein. Additionally, ESMFold given a single sequence as input outperforms both AF2 and RoseTTAFold, and can compete closely with RoseTTAFold even when given full MSAs information. Prediction speed displayed by ESMFold was reported to be faster compared to the existing programs, which can help in addressing the ever-growing protein sequence information compared to lagging growth of the structural database. They concluded that the LM employed for ESMFold is able to learn information similar to AF2 with MSAs data and that LM has a significant contribution to atomic-resolution structure prediction performance on rare proteins.

Diving into the complex quaternary structures of proteins, protein complexes such as antibodies are produced as an immune response to invading pathogens. Antibodies are made up of two pairs of heavy (one variable domain and three constant domains) and light chains (one variable domain and constant domain), of which three loops can be found respectively on each of the respective variable domains (L1, L2, L3, and H1, H2, H3) (Graves et al. 2020). These variable domains form complementarity-determining regions (CDRs) that are crucial in determining specific antibodies binding activity (Akbar et al. 2021, 2022; Polonelli et al. 2008). Efforts in antibody studies have led to the discovery of a canonical set of structural conformation displayed for five of the six CDRs (L1, L2, L3, H1, and H2). In contrast, this was not observed for CDR H3 which was highly variable in length and amino acid sequences (Teplyakov et al. 2016). As antibodies represent a unique group of proteins, development of an antibody language model (ALM) for prediction would definitely outperform a trained protein language model that covers a holistic range of protein.

ALM AbLang outperformed both IMGT germlines and protein language model ESM-1b in terms of a faster completion time and capability in restoring missing residues of antibody sequences (Olsen, Moal & Deane 2022). In AbLang, two separate antibody models were trained, one for heavy and another for light chains, with an imbalanced data set of 14 million heavy sequences and 187 thousand light sequences retrieved from the Observed Antibody Space (OAS) database. Olsen and colleagues designed the program to be able to generate three different useful representations of the antibody sequences, namely (i) res-codings useful for residue specific predictions, (ii) seq-codings useful for sequence specific predictions, and (iii) generating likelihood of amino acids at their respective specific positions in a given antibody sequence, which is handy for antibody engineering. Consequently, when comparing the performance of information extraction on B-cell sequence between the ALM against ESM-1b, AbLang was able to scrutinize better by separating the sequences based on V-genes into smaller clusters. Further, a clearer distinct segregation between naïve and memory B-cells was successfully achieved by AbLang. Within the OAS database, more than 40% of the sequences identified have 15 missing residues at the N-terminal. Utilizing the seq-codings representation information generated as a means to restore missing residues, AbLang displayed very similar performance with the IMGT germlines without needing any additional germlines information.

Another antibody LM, Antibody-specific Bidirectional Encoder Representation from Transformers (AntiBERTa), was reported to outperform two existing PLMs (ProtBert and Sapiens) and exhibited better B cell receptors representation when compared to ProtBERT that was assigned with a smaller dataset (Choi 2022; Leem et al. 2022). The authors employed a 12-layer transformer model to train on 57 million human BCR sequences (biased data sets of 42 million heavy and 15 million light chains), based on the RoBERTa architecture that allows a more direct comparison to be established (Liu et al. 2019). Performance validation reported capabilities of AntiBERTa in better distinguishing naïve and memory B-cells than the two PLMs. Integrated with a self-attention mechanism that gives informational embedding for each amino acid in the BCR sequence, AntiBERTa is more invested in what is functionally important for specific binding compared to ProtBERT finding a conserved disulfide bridge for all antibodies. When compared in the efficiency in paratope prediction against other existing paratope prediction approaches (Parapred, ProABC-2, ProtBERT, and Sapiens), AntiBERTa surpassed all of them. The authors described self-attention changes as the element of AntiBERTa allowed it to correctly predict paratope positions of both CDR and non-CDR positions. Interested readers are referred to the design methods of a linguistic-based formalization of the antibody language (Vu et al. 2022)

In summary, advances in cancer, computer vision, and protein language research have been primarily driven by the accumulation of large training datasets and the development of highly sophisticated deep learning architectures. Typically large attention-based models are trained on datasets in the order of $10^6$–$10^7$ data points. In contrast to NTD research where datasets remain restricted at the scale of $10^2$–$10^3$ (up to five orders of magnitude lower), shallow machine learning methods, namely RF, SVM, LR, among others, are more prevalent. The lack of large datasets restrict the widespread application of large deep learning models for the discovery of new NTD therapeutics and thus hamper the potential for efficient management and eradication of these diseases.

## On regional collaboration, data, and infrastructure sharing

The Southeast Asian countries are strategically located and exceptionally diverse in culture. Apart from the geographic proximity of Member States of the ASEAN regional association, the countries share a few other similarities of having densely populated communities, mineral-rich economies that open throughout the globe, and share similar tropical and subtropical climates. Hence, when there's a disease outbreak reported among any of the SEA countries, chances of the imported disease to neighboring countries are very high. Hence, there is a need to circumvent the matter through regional collaboration, and data plus infrastructure sharing among the SEA countries.

As previously described, the SEA region is endemic to vector-borne diseases such as arboviral diseases (dengue, LF, and malaria), leptospirosis, cysticercosis, and rabies. These diseases require up-to-date, robust, and comprehensive information on presence, species-strain diversity, ecology, environmental and geographical information regarding the organisms that carry and transmit the infectious agents. As such, the Malaria Atlas Project (MAP) (https://malariaatlas.org/) is an open-access database and WHO collaborating platform for geospatial disease modeling to project spatial limits, prevalence and endemicity of malaria in all locations around the world. The European Centre for Disease Prevention and Control (ECDC) (https://www.ecdc.europa.eu/) is an open-access database on dengue surveillance, threats, and outbreaks governed by the European Union. The database consists of almost all NTDs and other diseases that are of public health concern. The Global Atlas of Helminth Infections (GAHI) (https://www.thiswormyworld.org/) is an open-access database containing geographical distribution of neglected tropical diseases transmitted by worms: soil-transmitted helminthiasis, schistosomiasis, and lymphatic filariasis. All GAHI resources are available on an open access basis but up till the year 2015 only.

Another publicly available but somewhat geographically irrelevant to SEA region is the WHO-driven Expanded Special Project for the Elimination of NTDs (ESPEN) (https://espen.afro.who.int/). However, the ESPEN portal only contains survey data sets of NTDs in Africa. In response to the neglect of melioidosis, Melioidosis.info (https://www.melioidosis.info/infobox.aspx?pageID=101) serves as an online-platform for reporting melioidosis cases and for disseminating information of melioidosis for public, researchers and health policy makers. Two other notable and frequently visited database for health metrics and disease related data retrievals that have have been actively mentioned throughout this review are none other than WHO's Global Health Observatory (GHO) (https://www.who.int/data/gho) and Global Health Data Exchange (GHDx) (https://ghdx.healthdata.org/) , where both the Global Health Estimate (GHE) 2019 data and Global Burden of Disease Study (GBD) 2019 could be retrieved respectively by interested readers.

Global Alliance for Rabies Control (GARC) (https://rabiesalliance.org/) is the leading international rabies non-profit organization set to work with international stakeholders, governments and local partners to raise awareness about rabies, encourage collaboration, and build the evidence needed to increase political commitment and funding to end dog rabies in every country. Their main team of nine members work across three established work networks of ARACON (Asian Rabies Control Network), MERACON (Middle East, Eastern Europe, Central Asia and North Africa Rabies Control Network), and PARACON (Pan-African Rabies Control Network) as an effort to end rabies. International body responsible for infrastructure sharing in combating LF is the Global Alliance to Eliminate Lymphatic Filariasis (GAELF) (https://www.gaelf.org/). GAELF is a steering body aimed at bringing relevant partners to support the GPELF established by WHO via political, financial and technical resources mobilization. In response to the outbreak of leptospirosis, Global Leptospirosis Environment Action Network (GLEAN) (https://sites.google.com/site/gleanlepto/home?authuser=0) was created to reduce the global impact of leptospirosis through better understanding of the relationship between its occurrence and various associated factors including environmental, biological, ecological, economic and demographic factors, providing more timely warnings of outbreaks and identifying prevention and control strategies.

Data and infrastructure sharing is undeniably crucial for NTDs since the scale of publicly available datasets for NTDs research is dwarfed by other fields such as cancer, computer vision, and protein/antibody research. Efforts displayed by each governing body in maintaining and keeping up-to-date open-access databases or infrastructure and technical outreach organizations are to ensure that every country would have the latest disease intelligence and technical skills in order for effective surveillance, preventive, and disease management control to be executed according to each country's governing leadership. Importance of having centralized data sharing at a regional-scale has been highlighted in a study by Alemu et al. (2022). With access to publicly available standardized survey and treatment coverage data, which was at first unavailable probably due to absence of reports by the country's Ministry of Health to WHO, they were now able to access ample amounts of collected evidence pointing to the advantages of school-based deworming programs and LF MDA campaigns.

## Concluding remarks

NTDs impact nearly 2 billion people especially in countries with developing economies such as countries in the SEA and Western Pacific region causing reduction in productivity and substantial accumulation of Disability-Adjusted Life Years, DALY. Machine learning has been widely applied in fields such as cancer research, computer vision, and protein (language) modeling, however, the application of machine learning in NTDs research is hampered by the limited amount of data, the absence of centralized/standardized collaborative framework and the general lack of attention from public and private stakeholders alike. To unleash the full potential of machine learning to elevate the state-of-the-art NTDs surveillance, management, and treatment, increased investment in terms of research funding, public-private collaborative initiative, data accumulation and sharing are desperately needed.

## Author Contributions

CYK gathered the data, performed the analyses, and wrote the manuscript. RA wrote the manuscript and supervised the work. NMA wrote the manuscript and jointly supervised the work.

## Competing Interests

No competing interest declared.

## Grant Information

The authors acknowledge the Ministry of Higher Education, Malaysia for the financial support through Fundamental Research Grant Scheme (FRGS) funding (FRGS/1/2019/STG05/UKM/03/1) awarded to Norfarhan Mohd-Assaad. The APC was partially funded by Universiti Kebangsaan Malaysia (GGPM-2019-042).

# Acknowledgements

No acknowledgement declared.

# References

Abdullah, M., Kadivella, M., Sharma, R., Faisal, S.M. & Azam, S. 2021. Designing of multiepitope-based vaccine against Leptospirosis using Immuno-Informatics approaches. *bioRxiv*: 2021.02.22.431920.

Abdullah, N.M., Mohammad, W.M.Z.W., Shafei, M.N., Sukeri, S., Idris, Z., Arifin, W.N., Nozmi, N., Saudi, S.N.S., Samsudin, S., Zainudin, A.W., Hamat, R.A., Ibrahim, R., Masri, S.N., Saliluddin, S.M., Daud, A., Osman, M. & Jamaluddin, T.Z.M.T. 2019. Leptospirosis and its prevention: Knowledge, attitude and practice of urban community in Selangor, Malaysia. *BMC Public Health* 19(1): 1–8.

Abela-Ridder, B., Biswas, G., Mbabazi, P.S., Craven, M., Gerber, A., Hartenstein, L., Vlcek, J., Malecela, M.N., Polo, M.R., Tiendrebeogo, A., Castellanos, L., Nicholls, S., Jamsheed, M., Lin, Z., Gasimov, E., Atta, H. & Warusavithana, S. 2020. Ending the neglect to attain the sustainable development goals: a road map for neglected tropical diseases 2021–2030. *Who*, hlm.

Afolabi, M.O., Adebiyi, A., Cano, J., Sartorius, B., Greenwood, B., Johnson, O. & Wariri, O. 2022. Prevalence and distribution pattern of malaria and soil-transmitted helminth co-endemicity in sub-Saharan Africa, 2000–2018: A geospatial analysis. *PLOS Neglected Tropical Diseases* 16(9): e0010321.

Aguilera-Pesantes, D., Robayo, L.E., Méndez, P.E., Mollocana, D., Marrero-Ponce, Y., Torres, F.J. & Méndez, M.A. 2017. Discovering key residues of dengue virus NS2b-NS3-protease: New binding sites for antiviral inhibitors design. *Biochemical and Biophysical Research Communications* 492(4): 631–642.

Akbar, R., Bashour, H., Rawat, P., Robert, P.A., Smorodina, E., Cotet, T.S., Flem-Karlsen, K., Frank, R., Mehta, B.B., Vu, M.H., Zengin, T., Gutierrez-Marcos, J., Lund-Johansen, F., Andersen, J.T. & Greiff, V. 2022. Progress and challenges for the machine learning-based design of fit-for-purpose monoclonal antibodies. *mAbs* 14(1).

Akbar, R., Robert, P.A., Safonova, Y., Sandve, G.K. & Greiff, V. 2021. Article A compact vocabulary of paratope-epitope interactions enables predictability of antibody- antigen binding ll ll A compact vocabulary of paratope-epitope interactions enables predictability of antibody-antigen binding.

Akinsanya B, Adubi Taiwo, Macauley Adedamola, O.C. 2021. An investigation on the epidemiology and risk factors associated with soil-transmitted helminth infections in Ijebu East Local Government Area, Ogun State, Nigeria. *Scientific African* 12: e00757.

Alelign, T., Degarege, A. & Erko, B. 2015. Soil-Transmitted Helminth Infections and Associated Risk Factors among Schoolchildren in Durbete Town, Northwestern Ethiopia. *Journal of Parasitology Research* 2015(March 2010): 1–6.

Alemu, Y., Degefa, T., Bajiro, M. & Teshome, G. 2022. Prevalence and intensity of soil-transmitted helminths infection among individuals in model and non-model households, South West Ethiopia: A comparative cross-sectional community based study. *PLoS one* 17(10): e0276137.

Alley, E.C., Khimulya, G., Biswas, S., AlQuraishi, M. & Church, G.M. 2019. Unified rational protein engineering with sequence-based deep representation learning. *Nature Methods* 16(12): 1315–1322.

Almeida, D.S., Paz, L.N., De Oliveira, D.S., Silva, D.N., Ristow, P., Hamond, C., Costa, F., Portela, R.W., Estrela-Lima, A. & Pinna, M.H. 2019. Investigation of chronic infection by Leptospira spp. In asymptomatic sheep slaughtered in slaughterhouse. *PLoS ONE* 14(5): 1–13.

Alqahtani, A. 2022. Application of Artificial Intelligence in Discovery and Development of Anticancer and Antidiabetic Therapeutic Agents. *Evidence-based Complementary and Alternative Medicine* 2022.

Andre-Fontaine, G., Aviat, F. & Thorin, C. 2015. Waterborne Leptospirosis: Survival and Preservation of the Virulence of Pathogenic Leptospira spp. in Fresh Water. *Current Microbiology* 71(1): 136–142.

Ardila, D., Kiraly, A.P., Bharadwaj, S., Choi, B., Reicher, J.J., Peng, L., Tse, D., Etemadi, M., Ye, W., Corrado, G., Naidich, D.P. & Shetty, S. 2019. End-to-end lung cancer screening with three-dimensional deep learning on low-dose chest computed tomography. *Nature Medicine* 25(6): 954–961.

Baek, M., DiMaio, F., Anishchenko, I., Dauparas, J., Ovchinnikov, S., Lee, G.R., Wang, J., Cong, Q., Kinch, L.N., Schaeffer, R.D., Millán, C., Park, H., Adams, C., Glassman, C.R., DeGiovanni, A., Pereira, J.H., Rodrigues, A. V., van Dijk, A.A., Ebrecht, A.C., Opperman, D.J., Sagmeister, T., Buhlheller, C., Pavkov-Keller, T., Rathinaswamy, M.K., Dalwadi, U., Yip, C.K., Burke, J.E., Garcia, K.C., Grishin, N. V., Adams, P.D., Read, R.J. & Baker, D. 2021. Accurate prediction of protein structures and interactions using a three-track neural network. *Science* 373(6557): 871–876.

Bamaiyi, P.H. 2015. 2015 Outbreak of Canine Rabies in Malaysia: Review, Analysis and Perspectives. *Journal of Veterinary Advances* 5(12): 1181.

Becker, S.L., Liwanag, H.J., Snyder, J.S., Akogun, O., Belizario, V., Freeman, M.C., Gyorkos, T.W., Imtiaz, R., Keiser, J., Krolewiecki, A., Levecke, B., Mwandawiro, C., Pullan, R.L., Addiss, D.G. & Utzinger, J. 2018. Toward the 2020 goal of soil-transmitted helminthiasis control and elimination. *PLoS Neglected Tropical Diseases* 12(8): 1–17.

Bepler, T. & Berger, B. 2021. Learning the protein language: Evolution, structure, and function. *Cell Systems* 12(6): 654-669.e3.

Bertsimas, D. & Wiberg, H. 2020. Machine Learning in Oncology: Methods, Applications, and Challenges. *JCO Clinical Cancer Informatics*(4): 885–894.

Bethony, J., Brooker, S., Albonico, M., Geiger, S.M., Loukas, A., Diemert, D. & Hotez, P.J. 2006. Soil-transmitted helminth infections: ascariasis, trichuriasis, and hookworm. *Lancet* 367(9521): 1521–1532.

Bhadoria, P., Gupta, G. & Agarwal, A. 2021. Viral pandemics in the past two decades: An overview. *Journal of Family Medicine and Primary Care* 10(8): 2745.

Birnie, E., Virk, H.S., Savelkoel, J., Spijker, R., Bertherat, E., Dance, D.A.B., Limmathurotsakul, D., Devleesschauwer, B., Haagsma, J.A. & Wiersinga, W.J. 2019. Global burden of melioidosis in 2015: a systematic review and data synthesis. *The Lancet Infectious Diseases* 19(8): 892–902.

Borlee, G.I., Plumley, B.A., Martin, K.H., Somprasong, N., Mangalea, M.R., Islam, M.N., Burtnick, M.N., Brett, P.J., Steinmetz, I., AuCoin, D.P., Belisle, J.T., Crick, D.C., Schweizer, H.P. & Borlee, B.R. 2017. Genome-scale analysis of the genes that contribute to Burkholderia pseudomallei biofilm formation identifies a crucial exopolysaccharide biosynthesis gene cluster. *PLoS Neglected Tropical Diseases* 11(6): 1–29.

Braack, L., Gouveia De Almeida, A.P., Cornel, A.J., Swanepoel, R. & De Jager, C. 2018. Mosquito-borne arboviruses of African origin: Review of key viruses and vectors. *Parasites and Vectors* 11(1).

Brooker, S., Clements, A.C.A. & Bundy, D.A.P. 2006. Global Epidemiology, Ecology and Control of Soil-Transmitted Helminth Infections. *International Journal of Applied Mathematics and Statistics*, hlm. 221–261.

Brown, C.M., Slavinski, S., Ettestad, P., Sidwa, T.J. & Sorhage, F.E. 2016. Compendium of animal rabies prevention and control, 2016. *Journal of the American Veterinary Medical Association* 248(5): 505–517.

Castro, G.A. 1996. Helminths: Structure, Classification, Growth, and Development. Dlm. Baron, S. (pnyt.). hlm. Galveston (TX).

De Castro Poncio, L., Dos Anjos, F.A., De Oliveira, D.A., Rebechi, D., De Oliveira, R.N., Chitolina, R.F., Fermino, M.L., Bernardes, L.G., Guimarães, D., Lemos, P.A., Silva, M.N.E., Silvestre, R.G.M., Bernardes, E.S. & Paldi, N. 2021. Novel Sterile Insect Technology Program Results in Suppression of a Field Mosquito Population and Subsequently to Reduced Incidence of Dengue. *Journal of Infectious Diseases* 224(6): 1005–1014.

Casulli, A. 2021. New global targets for ntds in the who roadmap 2021–2030. *PLoS Neglected Tropical Diseases* 15(5): 1–10.

Chakravorty, A. & Heath, C.H. 2019. Melioidosis 48(5): 327–332.

Chaowagul, W., White, N.J., Dance, D.A.B., Wattanagoon, Y., Naigowit, P., Davis, T.M.E., Looareesuwan, S. & Pitakwatchara, N. 1989. Melioidosis: A Major Cause of Community-Acquired Septicemia in Northeastern Thailand. *Journal of Infectious Diseases* 159(5): 890–899.

Charoenkwan, P., Schaduangrat, N., Lio, P., Moni, M.A., Chumnanpuen, P. & Shoombuatong, W. 2022. iAMAP-SCM: A Novel Computational Tool for Large-Scale Identification of Antimalarial Peptides Using Estimated Propensity Scores of Dipeptides. *ACS Omega*.

Cheng, A.C. & Currie, B.J. 2005. Melioidosis: Epidemiology, Pathophysiology, and Management. *Clinical Microbiology Reviews* 18(2): 383–416.

Cheng, A.C., Jacups, S.P., Ward, L. & Currie, B.J. 2008. Melioidosis and Aboriginal seasons in northern Australia. *Transactions of the Royal Society of Tropical Medicine and Hygiene* 102: S26–S29.

Chong, N.S., Hardwick, R.J., Smith, S.R., Truscott, J.E. & Anderson, R.M. 2022. A prevalence-based transmission model for the study of the epidemiology and control of soil-transmitted helminthiasis. *PLoS ONE* 17(8 August): 1–28.

Chowdhury, R., Bouatta, N., Biswas, S., Rochereau, C., Church, G.M., Sorger, P.K. & Alquraishi, M. 2021. Single-sequence protein structure prediction using language models from deep learning. *AIChE Annual Meeting, Conference Proceedings* 2021-Novem.

Christophers, S.R. 1951. History of Malaria. *British Medical Journal* 1(4711): 865–866.

Condori, R.E., Niezgoda, M., Lopez, G., Matos, C.A., Mateo, E.D., Gigante, C., Hartloge, C., Filpo, A.P., Haim, J., Satheshkumar, P.S., Petersen, B., Wallace, R., Olson, V. & Li, Y. 2020. Using the LN34 pan-lyssavirus real-time RT-PCR assay for rabies diagnosis and rapid genetic typing from formalin-fixed human brain tissue. *Viruses* 12(1).

Costa, F., Hagan, J.E., Calcagno, J., Kane, M., Torgerson, P., Martinez-Silveira, M.S., Stein, C., Abela-Ridder, B. & Ko, A.I. 2015. Global Morbidity and Mortality of Leptospirosis: A Systematic Review. *PLoS Neglected Tropical Diseases* 9(9): 0–1.

Currie, B.J. 2015. Melioidosis: Evolving concepts in epidemiology, pathogenesis, and treatment. *Seminars in Respiratory and Critical Care Medicine* 36(1): 111–125.

Dadang, S.A. 2019. an Overview of Rabies Outbreaks in Malaysia, Ordinances and Laws. *Malaysian Journal of Veterinary Research* 10(2): 148–158.

Dara, S., Dhamercherla, S., Jadav, S.S., Babu, C.M. & Ahsan, M.J. 2022. Machine Learning in Drug Discovery: A Review. *Artificial Intelligence Review*, hlm. Springer Netherlands.:

Davenport, T. & Kalakota, R. 2019. The potential for artificial intelligence in healthcare. *Future Healthcare Journal* 6(2): 94–98.

David, A., Islam, S., Tankhilevich, E. & Sternberg, M.J.E. 2022. The AlphaFold Database of Protein Structures: A Biologist's Guide. *Journal of Molecular Biology* 434(2): 167336.

Dickson, B.F.R., Graves, P.M. & McBride, W.J. 2017. Lymphatic filariasis in mainland Southeast Asia: A systematic review and meta-analysis of prevalence and disease burden. *Tropical Medicine and Infectious Disease* 2(3).

Division, D.C. 2018. Interim Guideline for Human Rabies Prevention & Control in Malaysia. *Interim Guideline for Prevention and Control of Human Rabies in Malaysia*.

Dobigny, G., Gauthier, P., Houéménou, G., Choplin, A., Dossou, H.-J., Badou, S., Etougbétché, J., Bourhy, P., Koffi, S., Durski, K., Bertherat, E. & Picardeau, M. 2018. Leptospirosis and Extensive Urbanization in West Africa: A Neglected and Underestimated Threat? *Urban Science* 2(2): 29.

Douglass, J., Graves, P., Lindsay, D., Becker, L., Roineau, M., Masson, J., Aye, N.N., Win, S.S., Wai, T., Win, Y.Y. & Gordon, S. 2017. Lymphatic filariasis increases tissue compressibility and extracellular fluid in lower limbs of asymptomatic young people in central Myanmar. *Tropical Medicine and Infectious Disease* 2(4): 1–14.

Elnaggar, A., Heinzinger, M., Dallago, C., Rehawi, G., Wang, Y., Jones, L., Gibbs, T., Feher, T., Angerer, C., Steinegger, M., Bhowmik, D. & Rost, B. 2021. ProtTrans: Towards Cracking the Language of Lifes Code Through Self-Supervised Deep Learning and High Performance Computing. *IEEE Transactions on Pattern Analysis and Machine Intelligence*: 1–1.

Elumalai, E. 2022. Characterization and Prediction of Dengue Virus targeting peptides based on three class of descriptors using k-NN and Random Forest: 1–17.

Famakinde, D.O. 2018. Mosquitoes and the Lymphatic Filarial Parasites: Research Trends and Budding Roadmaps to Future Disease Eradication. *Tropical Medicine and Infectious Disease* 3(1).

Galipó, E., Dixon, M.A., Fronterrè, C., Cucunubá, Z.M., Basáñez, M.G., Stevens, K., Flórez Sánchez, A.C. & Walker, M. 2021. Spatial distribution and risk factors for human cysticercosis in Colombia. *Parasites and Vectors* 14(1): 1–15.

Galyov, E.E., Brett, P.J. & Deshazer, D. 2010. Molecular insights into Burkholderia pseudomallei and Burkholderia mallei pathogenesis. *Annual Review of Microbiology* 64: 495–517.

Garba, B., Bahaman, A.R., Bejo, S.K., Zakaria, Z., Mutalib, A.R. & Bande, F. 2018. Major epidemiological factors associated with leptospirosis in Malaysia. *Acta Tropica*

178(November 2017): 242–247.

García, H.H., Gonzalez, A.E., Evans, C.A.W. & Gilman, R.H. 2003. Taenia solium cysticercosis. *Lancet* 362(9383): 547–556.

Geoffrey, B., Sanker, A., Madaj, R., Tresanco, M.S.V., Upadhyay, M. & Gracia, J. 2022. A program to automate the discovery of drugs for West Nile and Dengue virus—programmatic screening of over a billion compounds on PubChem, generation of drug leads and automated in silico modelling. *Journal of Biomolecular Structure and Dynamics* 40(10): 4293–4300.

Guirakhoo, F., Arroyo, J., Pugachev, K. V., Miller, C., Zhang, Z.-X., Weltzin, R., Georgakopoulos, K., Catalan, J., Ocran, S., Soike, K., Ratterree, M. & Monath, T.P. 2001. Construction, Safety, and Immunogenicity in Nonhuman Primates of a Chimeric Yellow Fever-Dengue Virus Tetravalent Vaccine. *Journal of Virology* 75(16): 7290–7304.

Haake, D.A. & Levett, P.N. 2015. Leptospirosis in Humans. *Journal of Biological Education*, hlm. 65–97.

Hatcher, C.L., Muruato, L.A. & Torres, A.G. 2015. Recent Advances in Burkholderia mallei and B. pseudomallei Research. *Current Tropical Medicine Reports* 2(2): 62–69.

Heinzinger, M., Elnaggar, A., Wang, Y., Dallago, C., Nechaev, D., Matthes, F. & Rost, B. 2019. Modeling aspects of the language of life through transfer-learning protein sequences. *BMC Bioinformatics* 20(1): 1–17.

Horwood, P.F., Tarantola, A., Goarant, C., Matsui, M., Klement, E., Umezaki, M., Navarro, S. & Greenhill, A.R. 2019. Health Challenges of the Pacific Region: Insights From History, Geography, Social Determinants, Genetics, and the Microbiome. *Frontiers in Immunology* 10(September): 1–16.

Hotez, P.J. & Lo, N.C. 2020. Neglected Tropical Diseases: Public Health Control Programs and Mass Drug Administration. *Hunter's Tropical Medicine and Emerging Infectious Diseases*, hlm. Tenth Edit. Elsevier Inc.:

Hugo, L.E., Rašić, G., Maynard, A.J., Ambrose, L., Liddington, C., Thomas, C.J.E., Nath, N.S., Graham, M., Winterford, C., Wimalasiri-Yapa, B.M.C.R., Xi, Z., Beebe, N.W. & Devine, G.J. 2022. Wolbachia wAlbB inhibit dengue and Zika infection in the mosquito Aedes aegypti with an Australian background. *bioRxiv*: 2022.03.22.485408.

Inusa, B.P.D., Hsu, L.L., Kohli, N., Patel, A., Ominu-Evbota, K., Anie, K.A. & Atoyebi, W. 2019. Sickle cell disease—genetics, pathophysiology, clinical presentation and treatment. *International Journal of Neonatal Screening* 5(2).

Jainul Fathima, A., Revathy, R., Balamurali, S. & Murugaboopathi, G. 2019. Prediction of Dengue-Human Protein Interaction Using Artificial Neural Network for Anti-Viral Drug Discovery. *SSRN Electronic Journal*(January).

Jannin, J., Solano, P., Quick, I. & Debre, P. 2017. The francophone network on neglected tropical diseases. *PLoS Neglected Tropical Diseases* 11(8): 2–6.

Jansen, F., Dorny, P., Gabriël, S., Dermauw, V., Johansen, M.V. & Trevisan, C. 2021. The survival and dispersal of Taenia eggs in the environment: what are the implications for transmission? A systematic review. *Parasites and Vectors* 14(1): 1–16.

Jumper, J., Evans, R., Pritzel, A., Green, T., Figurnov, M., Ronneberger, O., Tunyasuvunakool, K., Bates, R., Žídek, A., Potapenko, A., Bridgland, A., Meyer, C., Kohl, S.A.A., Ballard, A.J., Cowie, A., Romera-Paredes, B., Nikolov, S., Jain, R., Adler, J., Back, T., Petersen, S., Reiman, D., Clancy, E., Zielinski, M., Steinegger, M., Pacholska, M., Berghammer, T., Bodenstein, S., Silver, D., Vinyals, O., Senior, A.W., Kavukcuoglu, K., Kohli, P. & Hassabis, D. 2021. Highly accurate protein structure prediction with AlphaFold. *Nature* 596(7873): 583–589.

Khalid, A.Q., Rao Avupati, V. & Hussain, H. 2020. Machine Learning Model For Predicting Anti-Dengue Drugs: A Three-Dimensional Quantitative Structure-Activity Relationship (3D QSAR) Study. *International Journal Of Science & Technology Research* 9(6): 1107–1115.

Khan, S. & Vihinen, M. 2007. Spectrum of disease-causing mutations in protein secondary structures. *BMC Structural Biology* 7: 1–18.

Kourou, K., Exarchos, K.P., Papaloukas, C., Sakaloglou, P., Exarchos, T. & Fotiadis, D.I. 2021. Applied machine learning in cancer research: A systematic review for patient diagnosis, classification and prognosis. *Computational and Structural Biotechnology Journal* 19: 5546–5555.

Lalremruata, A., Jeyaraj, S., Engleitner, T., Joanny, F., Lang, A., Bélard, S., Mombo-Ngoma, G., Ramharter, M., Kremsner, P.G., Mordmüller, B. & Held, J. 2017. Species and genotype diversity of Plasmodium in malaria patients from Gabon analysed by next generation sequencing. *Malaria Journal* 16(1): 1–11.

Laranjo-González, M., Devleesschauwer, B., Trevisan, C., Allepuz, A., Sotiraki, S., Abraham, A., Afonso, M.B., Blocher, J., Cardoso, L., Correia Da Costa, J.M., Dorny, P., Gabriël, S., Gomes, J., Gómez-Morales, M.Á., Jokelainen, P., Kaminski, M., Krt, B., Magnussen, P., Robertson, L.J., Schmidt, V., Schmutzhard, E., Smit, G.S.A., Šoba, B., Stensvold, C.R., Starič, J., Troell, K., Rataj, A.V., Vieira-Pinto, M., Vilhena, M., Wardrop, N.A., Winkler, A.S. & Dermauw, V. 2017. Epidemiology of taeniosis/cysticercosis in Europe, a systematic review: Western Europe. *Parasites and Vectors* 10(1): 1–14.

Larsen, J.C. & Johnson, N.H. 2009. Pathogenesis of Burkholderia pseudomallei and Burkholderia mallei. *Military medicine* 174(6): 647–51.

Larson, B. 2019. Origin of Two Most Virulent Agents of Human Malaria: Plasmodium falciparum and Plasmodium vivax. *Malaria*, hlm. 13. IntechOpen.:

Lecun, Y., Bengio, Y. & Hinton, G. 2015. Deep learning. *Nature* 521(7553): 436–444.

Levett, P.N. 2015. Systematics of leptospiraceae. *Current Topics in Microbiology and Immunology* 387: 11–20.

Limmathurotsakul, D., Golding, N., Dance, D.A.B., Messina, J.P., Pigott, D.M., Moyes, C.L., Rolim, D.B., Bertherat, E., Day, N.P.J., Peacock, S.J. & Hay, S.I. 2016. Predicted global distribution of Burkholderia pseudomallei and burden of melioidosis. *Nature Microbiology* 1(1): 6–10.

Limmathurotsakul, D., Wuthiekanun, V., Chierakul, W., Cheng, A.C., Maharjan, B., Chaowagul, W., White, N.J., Day, N.P.J. & Peacock, S.J. 2005. Role and Significance of Quantitative Urine Cultures in Diagnosis of Melioidosis 43(5): 2274–2276.

Lin, Z., Akin, H., Rao, R., Hie, B., Zhu, Z., Lu, W., Costa, A. dos S., Fazel-Zarandi, M., Sercu, T., Candido, S. & Rives, A. 2022. Language models of protein sequences at the scale of evolution enable accurate structure prediction. *bioRxiv*: 2022.07.20.500902.

Littmann, M., Heinzinger, M., Dallago, C., Olenyi, T. & Rost, B. 2021. Embeddings from deep learning transfer GO annotations beyond homology. *Scientific Reports* 11(1): 1–14.

Liu, W., Li, Y., Shaw, K.S., Learn, G.H., Plenderleith, L.J., Malenke, J.A., Sundararaman, S.A., Ramirez, M.A., Crystal, P.A., Smith, A.G., Bibollet-Ruche, F., Ayouba, A., Locatelli, S., Esteban, A., Mouacha, F., Guichet, E., Butel, C., Ahuka-Mundeke, S., Inogwabini, B.-I., Ndjango, J.-B.N., Speede, S., Sanz, C.M., Morgan, D.B., Gonder, M.K., Kranzusch, P.J., Walsh, P.D., Georgiev, A. V., Muller, M.N., Piel, A.K., Stewart, F.A., Wilson, M.L., Pusey, A.E., Cui, L., Wang, Z., Färnert, A., Sutherland, C.J., Nolder, D., Hart, J.A., Hart, T.B., Bertolani, P., Gillis, A., LeBreton, M., Tafon, B., Kiyang, J., Djoko, C.F., Schneider, B.S., Wolfe, N.D., Mpoudi-Ngole, E., Delaporte, E., Carter, R., Culleton, R.L., Shaw, G.M., Rayner, J.C., Peeters, M., Hahn, B.H. & Sharp, P.M. 2014. African origin of the malaria parasite Plasmodium vivax. *Nature Communications* 5(1): 3346.

Lourenço, C., Tatem, A.J., Atkinson, P.M., Cohen, J.M., Pindolia, D., Bhavnani, D. & Le Menach, A. 2019. Strengthening surveillance systems for malaria elimination: A global landscaping of system performance, 2015-2017. *Malaria Journal* 18(1): 1–11.

Lupenza, E.T., Gasarasi, D.B. & Minzi, O.M. 2022. Lymphatic filariasis elimination status: Wuchereria bancrofti infections in human populations and factors contributing to continued transmission after seven rounds of mass drug administration in Masasi District, Tanzania. *PLoS ONE* 17(1 January): 1–11.

Lupo, U., Sgarbossa, D. & Bitbol, A.F. 2022. Protein language models trained on multiple sequence alignments learn phylogenetic relationships. *Nature Communications* 13(1):

1–23.

Lustigman, S., Prichard, R.K., Gazzinelli, A., Grant, W.N., Boatin, B.A., McCarthy, J.S. & Basáñez, M.G. 2012. A research agenda for helminth diseases of humans: The problem of helminthiases. *PLoS Neglected Tropical Diseases* 6(4).

Manuel, M., Ramanujam, K. & Ajjampur, S.S.R. 2021. Molecular Tools for Diagnosis and Surveillance of Soil-Transmitted Helminths in Endemic Areas. *Parasitologia* 1(3): 105–118.

Marquet, C., Heinzinger, M., Olenyi, T., Dallago, C., Erckert, K., Bernhofer, M., Nechaev, D. & Rost, B. 2022. Embeddings from protein language models predict conservation and variant effects. *Human Genetics* 141(10): 1629–1647.

McComb, M., Bies, R. & Ramanathan, M. 2022. Machine learning in pharmacometrics: Opportunities and challenges. *British Journal of Clinical Pharmacology* 88(4): 1482–1499.

Mitra, A.K. & Mawson, A.R. 2017. Neglected tropical diseases: Epidemiology and global burden. *Tropical Medicine and Infectious Disease* 2(3).

Mogaji, H.O., Dedeke, G.A., Bada, B.S., Bankole, S., Adeniji, A., Fagbenro, M.T., Omitola, O.O., Oluwole, A.S., Odoemene, N.S., Abe, E.M., Mafiana, C.F. & Ekpo, U.F. 2020. Distribution of ascariasis, trichuriasis and hookworm infections in Ogun State, Southwestern Nigeria. *PLoS ONE* 15(6): 1–16.

Mogaji, H.O., Johnson, O.O., Adigun, A.B., Adekunle, O.N., Bankole, S., Dedeke, G.A., Bada, B.S. & Ekpo, U.F. 2022. Estimating the population at risk with soil transmitted helminthiasis and annual drug requirements for preventive chemotherapy in Ogun State, Nigeria. *Scientific Reports* 12(1): 1–12.

Molyneux, D. 2019. Neglected Tropical Diseases - East Asia. J. Utzinger, P. Yap, M. Bratschi & P. Steinmann (Pnyt.) *Community Eye Health Journal*, hlm. Springer International Publishing: Cham.

Montresor, A., Mupfasoni, D., Mikhailov, A., Mwinzi, P., Lucianez, A., Jamsheed, M., Gasimov, E., Warusavithana, S., Yajima, A., Bisoffi, Z., Buonfrate, D., Steinmann, P., Utzinger, J., Levecke, B., Vlaminck, J., Coolsid, P., Vercruysse, J., Cringoli, G., Rinaldi, L., Blouinid, B. & Gyorkos, T.W. 2020. The global progress of soil-transmitted helminthiases control in 2020 and world health organization targets for 2030. *PLoS Neglected Tropical Diseases* 14(8): 1–17.

Mswahili, M.E., Martin, G.L., Woo, J., Choi, G.J. & Jeong, Y.S. 2021. Antimalarial drug predictions using molecular descriptors and machine learning against plasmodium falciparum. *Biomolecules* 11(12): 1–15.

Muñoz-Antoli, C., Pérez, P., Pavón, A., Toledo, R. & Esteban, J.G. 2022. High intestinal parasite infection detected in children from Región Autónoma Atlántico Norte (R.A.A.N.) of Nicaragua. *Scientific Reports* 12(1): 1–10.

Ofer, D., Brandes, N. & Linial, M. 2021. The language of proteins: NLP, machine learning & protein sequences. *Computational and Structural Biotechnology Journal* 19: 1750–1758.

Oguike, O.E., Ugwuishiwu, C.H., Asogwa, C.N., Nnadi, C.O., Obonga, W.O. & Attama, A.A. 2022. Systematic review on the application of machine learning to quantitative structure–activity relationship modeling against Plasmodium falciparum. *Molecular Diversity* 26(6): 3447–3462.

Ong, E., Wang, H., Wong, M.U., Seetharaman, M., Valdez, N. & He, Y. 2020. Vaxign-ML: Supervised machine learning reverse vaccinology model for improved prediction of bacterial protective antigens. *Bioinformatics* 36(10): 3185–3191.

Oyewole, O.E. & Simon-Oke, I.A. 2022. Ecological risk factors of soil-transmitted helminths infections in Ifedore district, Southwest Nigeria. *Bulletin of the National Research Centre* 46(1).

Pal, M., Tewari, A., Gerbaba, N.D. & Shuramo, M.Y. 2022. Melioidosis : An emerging yet neglected bacterial zoonosis. *Journal of Bacteriology & Mycology* 10(2): 32–37.

Pearson, W.R. 2013. An Introduction to Sequence Similarity ("Homology") Searching. *Current Protocols in Bioinformatics* 42(1): 1286–1292.

Polonelli, L., Pontón, J., Elguezabal, N., Moragues, M.D., Casoli, C., Pilotti, E., Ronzi, P., Dobroff, A.S., Rodrigues, E.G., Juliano, M.A., Maffei, D.L., Magliani, W., Conti, S. & Travassos, L.R. 2008. Antibody complementarity-determining regions (CDRs) can display differential antimicrobial, antiviral and antitumor activities. *PLoS ONE* 3(6).

Ranathunge, T., Harishchandra, J., Maiga, H., Bouyer, J., Gunawardena, Y.I.N.S. & Hapugoda, M. 2022. Development of the Sterile Insect Technique to control the dengue vector Aedes aegypti (Linnaeus) in Sri Lanka. *PLoS ONE* 17(4 April): 1–15.

Rapin, N., Lund, O., Bernaschi, M. & Castiglione, F. 2010. Computational immunology meets bioinformatics: The use of prediction tools for molecular binding in the simulation of the immune system. *PLoS ONE* 5(4).

Rives, A., Meier, J., Sercu, T., Goyal, S., Lin, Z., Liu, J., Guo, D., Ott, M., Zitnick, C.L., Ma, J. & Fergus, R. 2021. Biological structure and function emerge from scaling unsupervised learning to 250 million protein sequences. *Proceedings of the National Academy of Sciences of the United States of America* 118(15).

Saba, N., Balwan, W.K. & Mushtaq, F. 2022. Burden of Malaria - A Journey Revisited. *Scholars

*Journal of Applied Medical Sciences* 10(6): 934–939.

Sarkar-Tyson, M. & Titball, R.W. 2009. Burkholderia mallei and Burkholderia pseudomallei. *Vaccines for Biodefense and Emerging and Neglected Diseases*, hlm. 831–843. Elsevier.:

Sataloff, R.T., Johns, M.M. & Kost, K.M. (n.d.). Rabies (Infection with Rabies Virus and other Lyssaviruses) [oie.int].

Sato, M.O., Adsakwattana, P., Fontanilla, I.K.C., Kobayashi, J., Sato, M., Pongvongsa, T., Fornillos, R.J.C. & Waikagul, J. 2019. Odds, challenges and new approaches in the control of helminthiasis, an Asian study. *Parasite Epidemiology and Control* 4: e00083.

Sato, S. 2021. Plasmodium—a brief introduction to the parasites causing human malaria and their basic biology. *Journal of Physiological Anthropology* 40(1): 1–13.

Schlottau, K., Freuling, C.M., Müller, T., Beer, M. & Hoffmann, B. 2017. Development of molecular confirmation tools for swift and easy rabies diagnostics: 1–13.

Seera, S. & Nagarajaram, H.A. 2021. Effect of Disease Causing Missense Mutations on Intrinsically Disordered Regions in Proteins. *Protein & Peptide Letters* 29(3): 254–267.

Selvarajoo, S., Liew, J.W.K., Chua, T.H., Tan, W., Zaki, R.A., Ngui, R., Sulaiman, W.Y.W., Ong, P.S. & Vythilingam, I. 2022. Dengue surveillance using gravid oviposition sticky (GOS) trap and dengue non-structural 1 (NS1) antigen test in Malaysia: randomized controlled trial. *Scientific Reports* 12(1): 1–12.

Shao, D., Dai, Y., Li, N., Cao, X., Zhao, W., Cheng, L., Rong, Z., Huang, L., Wang, Y. & Zhao, J. 2022. Artificial intelligence in clinical research of cancers. *Briefings in Bioinformatics* 23(1): 1–12.

Silber, S.A., Diro, E., Workneh, N., Mekonnen, Z., Levecke, B., Steinmann, P., Umulisa, I., Alemu, H., Baeten, B., Engelen, M., Hu, P., Friedman, A., Baseman, A. & Mrus, J. 2017. Efficacy and safety of a single-dose mebendazole 500 mg chewable, rapidly-disintegrating tablet for ascaris lumbricoides and trichuris trichiura infection treatment in pediatric patients: A double-blind, randomized, placebo-controlled, phase 3 study. *American Journal of Tropical Medicine and Hygiene* 97(6): 1851–1856.

Simo, F.B.N., Bigna, J.J., Kenmoe, S., Ndangang, M.S., Temfack, E., Moundipa, P.F. & Demanou, M. 2019. Dengue virus infection in people residing in Africa: a systematic review and meta-analysis of prevalence studies. *Scientific Reports* 9(1): 1–9.

Simon, D.W., Da Silva, Y.S., Zuccoli, G. & Clark, R.S.B. 2013. Acute Encephalitis. *Critical Care Clinics* 29(2): 259–277.

Singh, J., Arora, M.S., Sharma, S. & Shukla, J.B. 2022. Modeling the variable transmission rate

and various discharges on the spread of Malaria. *Electronic Research Archive* 31(1): 319–341.

Sridhar, S., Luedtke, A., Langevin, E., Zhu, M., Bonaparte, M., Machabert, T., Savarino, S., Zambrano, B., Moureau, A., Khromava, A., Moodie, Z., Westling, T., Mascareñas, C., Frago, C., Cortés, M., Chansinghakul, D., Noriega, F., Bouckenooghe, A., Chen, J., Ng, S.-P., Gilbert, P.B., Gurunathan, S. & DiazGranados, C.A. 2018. Effect of Dengue Serostatus on Dengue Vaccine Safety and Efficacy. *New England Journal of Medicine*: 327–340.

Srisawat, N., Thisyakorn, U., Ismail, Z., Rafiq, K. & Gubler, D.J. 2022. World Dengue Day: A call for action. *PLoS Neglected Tropical Diseases* 16(8): 2–10.

Stärk, H., Dallago, C., Heinzinger, M. & Rost, B. 2021. Light attention predicts protein location from the language of life. *Bioinformatics Advances* 1(1).

Sun, A.H., Liu, X.X. & Yan, J. 2020. Leptospirosis is an invasive infectious and systemic inflammatory disease. *Biomedical Journal* 43(1): 24–31.

Suratanee, A., Buaboocha, T. & Plaimas, K. 2021. Prediction of Human-Plasmodium vivax Protein Associations From Heterogeneous Network Structures Based on Machine-Learning Approach. *Bioinformatics and Biology Insights* 15.

Sykes, J.E., Reagan, K.L., Nally, J.E., Galloway, R.L. & Haake, D.A. 2022. Role of Diagnostics in Epidemiology, Management, Surveillance, and Control of Leptospirosis. *Pathogens* 11(4): 1–24.

Taylor, B. 2020. Artificial Intelligence in Oncology Drug Discovery and Development. *Artificial Intelligence in Oncology Drug Discovery and Development*, hlm.

The Center for Food Security and Public Health. 2012. Rabies and Lyssaviruses. *Disease data sheet*: 1–12.

Thomas, S.J. & Yoon, I.K. 2019. A review of Dengvaxia®: development to deployment. *Human Vaccines and Immunotherapeutics* 15(10): 2295–2314.

Tian, N., Zheng, J.X., Guo, Z.Y., Li, L.H., Xia, S., Lv, S. & Zhou, X.N. 2022. Dengue Incidence Trends and Its Burden in Major Endemic Regions from 1990 to 2019. *Tropical Medicine and Infectious Disease* 7(8).

Tsheten, T., Gray, D.J., Clements, A.C.A. & Wangdi, K. 2021. Epidemiology and challenges of dengue surveillance in the WHO South-East Asia Region. *Transactions of the Royal Society of Tropical Medicine and Hygiene* 115(6): 583–599.

Tsiang, H., De La Porte, S., Ambroise, D.J., Derer, M. & Koenig, J. 1986. Infection of cultured rat myotubes and neurons from the spinal cord by rabies virus. *Journal of Neuropathology* and Experimental Neurology 45(1): 28–42.

Tully, D. & Griffiths, C.L. 2021. Dengvaxia: the world's first vaccine for prevention of secondary dengue. *Therapeutic Advances in Vaccines and Immunotherapy* 9: 1–8.

Urbanskas, E., Karvelienė, B. & Radzijevskaja, J. 2022. Leptospirosis: classification, epidemiology, and methods of detection. A review. *Biologija* 68(2): 129–136.

Vinkeles Melchers, N.V.S., Stolk, W.A., van Loon, W., Pedrique, B., Bakker, R., Murdoch, M.E., de Vlas, S.J. & Coffeng, L.E. 2021. The burden of skin disease and eye disease due to onchocerciasis in countries formerly under the african programme for onchocerciasis control mandate for 1990, 2020, and 2030. *PLoS Neglected Tropical Diseases* 15(7): 1–18.

Vos, T., Lim, S.S., Abbafati, C., Abbas, K.M., Abbasi, M., Abbasifard, M., Abbasi-Kangevari, M., Abbastabar, H., Abd-Allah, F., Abdelalim, A., Abdollahi, M., Abdollahpour, I., Abolhassani, H., Aboyans, V., Abrams, E.M., Abreu, L.G., Abrigo, M.R.M., Abu-Raddad, L.J., Abushouk, A.I., Acebedo, A., Ackerman, I.N., Adabi, M., Adamu, A.A., Adebayo, O.M., Adekanmbi, V., Adelson, J.D., Adetokunboh, O.O., Adham, D., Afshari, M., Afshin, A., Agardh, E.E., Agarwal, G., Agesa, K.M., Aghaali, M., Aghamir, S.M.K., Agrawal, A., Ahmad, T., Ahmadi, A., Ahmadi, M., Ahmadieh, H., Ahmadpour, E., Akalu, T.Y., Akinyemi, R.O., Akinyemiju, T., Akombi, B., Al-Aly, Z., Alam, K., Alam, N., Alam, S., Alam, T., Alanzi, T.M., Albertson, S.B., Alcalde-Rabanal, J.E., Alema, N.M., Ali, M., Ali, S., Alicandro, G., Alijanzadeh, M., Alinia, C., Alipour, V., Aljunid, S.M., Alla, F., Allebeck, P., Almasi-Hashiani, A., Alonso, J., Al-Raddadi, R.M., Altirkawi, K.A., Alvis-Guzman, N., Alvis-Zakzuk, N.J., Amini, S., Amini-Rarani, M., Aminorroaya, A., Amiri, F., Amit, A.M.L., Amugsi, D.A., Amul, G.G.H., Anderlini, D., Andrei, C.L., Andrei, T., Anjomshoa, M., Ansari, F., Ansari, I., Ansari-Moghaddam, A., Antonio, C.A.T., Antony, C.M., Antriyandarti, E., Anvari, D., Anwer, R., Arabloo, J., Arab-Zozani, M., Aravkin, A.Y., Ariani, F., Ärnlöv, J., Aryal, K.K., Arzani, A., Asadi-Aliabadi, M., Asadi-Pooya, A.A., Asghari, B., Ashbaugh, C., Atnafu, D.D., Atre, S.R., Ausloos, F., Ausloos, M., Ayala Quintanilla, B.P., Ayano, G., Ayanore, M.A., Aynalem, Y.A., Azari, S., Azarian, G., Azene, Z.N., Babaee, E., Badawi, A., Bagherzadeh, M., Bakhshaei, M.H., Bakhtiari, A., Balakrishnan, S., Balalla, S., Balassyano, S., Banach, M., Banik, P.C., Bannick, M.S., Bante, A.B., Baraki, A.G., Barboza, M.A., Barker-Collo, S.L., Barthelemy, C.M., Barua, L., Barzegar, A., Basu, S., Baune, B.T., Bayati, M., Bazmandegan, G., Bedi, N., Beghi, E., Béjot, Y., Bello, A.K., Bender, R.G., Bennett, D.A., Bennitt, F.B., Bensenor, I.M., Benziger, C.P., Berhe, K., Bernabe, E., Bertolacci, G.J., Bhageerathy, R., Bhala, N., Bhandari, D., Bhardwaj, P., Bhattacharyya, K., Bhutta, Z.A., Bibi, S., Biehl, M.H., Bikbov, B., Bin Sayeed, M.S., Biondi, A., Birihane, B.M., Bisanzio, D., Bisignano, C., Biswas, R.K., Bohlouli, S., Bohluli, M., Bolla, S.R.R., Boloor, A., Boon-Dooley, A.S., Borges, G., Borzì, A.M., Bourne, R., Brady, O.J., Brauer, M., Brayne, C., Breitborde, N.J.K., Brenner, H., Briant, P.S., Briggs, A.M., Briko, N.I., Britton, G.B., Bryazka, D., Buchbinder, R., Bumgarner, B.R., Busse, R., Butt, Z.A., Caetano dos Santos, F.L., Cámera, L.L.A., Campos-Nonato, I.R., Car, J., Cárdenas, R., Carreras, G., Carrero, J.J., Carvalho, F., Castaldelli-Maia, J.M., Castañeda-Orjuela, C.A., Castelpietra, G., Castle, C.D., Castro, F., Catalá-López, F., Causey, K., Cederroth, C.R., Cercy, K.M., Cerin, E.,

Chandan, J.S., Chang, A.R., Charlson, F.J., Chattu, V.K., Chaturvedi, S., Chimed-Ochir, O., Chin, K.L., Cho, D.Y., Christensen, H., Chu, D.-T., Chung, M.T., Cicuttini, F.M., Ciobanu, L.G., Cirillo, M., Collins, E.L., Compton, K., Conti, S., Cortesi, P.A., Costa, V.M., Cousin, E., Cowden, R.G., Cowie, B.C., Cromwell, E.A., Cross, D.H., Crowe, C.S., Cruz, J.A., Cunningham, M., Dahlawi, S.M.A., Damiani, G., Dandona, L., Dandona, R., Darwesh, A.M., Daryani, A., Das, J.K., Das Gupta, R., das Neves, J., Dávila-Cervantes, C.A., Davletov, K., De Leo, D., Dean, F.E., DeCleene, N.K., Deen, A., Degenhardt, L., Dellavalle, R.P., Demeke, F.M., Demsie, D.G., Denova-Gutiérrez, E., Dereje, N.D., Dervenis, N., Desai, R., Desalew, A., Dessie, G.A., Dharmaratne, S.D., Dhungana, G.P., Dianatinasab, M., Diaz, D., Dibaji Forooshani, Z.S., Dingels, Z. V, Dirac, M.A., Djalalinia, S., Do, H.T., Dokova, K., Dorostkar, F., Doshi, C.P., Doshmangir, L., Douiri, A., Doxey, M.C., Driscoll, T.R., Dunachie, S.J., Duncan, B.B., Duraes, A.R., Eagan, A.W., Ebrahimi Kalan, M., Edvardsson, D., Ehrlich, J.R., El Nahas, N., El Sayed, I., El Tantawi, M., Elbarazi, I., Elgendy, I.Y., Elhabashy, H.R., El-Jaafary, S.I., Elyazar, I.R.F., Emamian, M.H., Emmons-Bell, S., Erskine, H.E., Eshrati, B., Eskandarieh, S., Esmaeilnejad, S., Esmaeilzadeh, F., Esteghamati, A., Estep, K., Etemadi, A., Etisso, A.E., Farahmand, M., Faraj, A., Fareed, M., Faridnia, R., Farinha, C.S. e S., Farioli, A., Faro, A., Faruque, M., Farzadfar, F., Fattahi, N., Fazlzadeh, M., Feigin, V.L., Feldman, R., Fereshtehnejad, S.-M., Fernandes, E., Ferrari, A.J., Ferreira, M.L., Filip, I., Fischer, F., Fisher, J.L., Fitzgerald, R., Flohr, C., Flor, L.S., Foigt, N.A., Folayan, M.O., Force, L.M., Fornari, C., Foroutan, M., Fox, J.T., Freitas, M., Fu, W., Fukumoto, T., Furtado, J.M., Gad, M.M., Gakidou, E., Galles, N.C., Gallus, S., Gamkrelidze, A., Garcia-Basteiro, A.L., Gardner, W.M., Geberemariyam, B.S., Gebrehiwot, A.M., Gebremedhin, K.B., Gebreslassie, A.A.A.A., Gershberg Hayoon, A., Gething, P.W., Ghadimi, M., Ghadiri, K., Ghafourifard, M., Ghajar, A., Ghamari, F., Ghashghaee, A., Ghiasvand, H., Ghith, N., Gholamian, A., Gilani, S.A., Gill, P.S., Gitimoghaddam, M., Giussani, G., Goli, S., Gomez, R.S., Gopalani, S.V., Gorini, G., Gorman, T.M., Gottlich, H.C., Goudarzi, H., Goulart, A.C., Goulart, B.N.G., Grada, A., Grivna, M., Grosso, G., Gubari, M.I.M., Gugnani, H.C., Guimaraes, A.L.S., Guimarães, R.A., Guled, R.A., Guo, G., Guo, Y., Gupta, R., Haagsma, J.A., Haddock, B., Hafezi-Nejad, N., Hafiz, A., Hagins, H., Haile, L.M., Hall, B.J., Halvaei, I., Hamadeh, R.R., Hamagharib Abdullah, K., Hamilton, E.B., Han, C., Han, H., Hankey, G.J., Haro, J.M., Harvey, J.D., Hasaballah, A.I., Hasanzadeh, A., Hashemian, M., Hassanipour, S., Hassankhani, H., Havmoeller, R.J., Hay, R.J., Hay, S.I., Hayat, K., Heidari, B., Heidari, G., Heidari-Soureshjani, R., Hendrie, D., Henrikson, H.J., Henry, N.J., Herteliu, C., Heydarpour, F., Hird, T.R., Hoek, H.W., Hole, M.K., Holla, R., Hoogar, P., Hosgood, H.D., Hosseinzadeh, M., Hostiuc, M., Hostiuc, S., Househ, M., Hoy, D.G., Hsairi, M., Hsieh, V.C., Hu, G., Huda, T.M., Hugo, F.N., Huynh, C.K., Hwang, B.-F., Iannucci, V.C., Ibitoye, S.E., Ikuta, K.S., Ilesanmi, O.S., Ilic, I.M., Ilic, M.D., Inbaraj, L.R., Ippolito, H., Irvani, S.S.N., Islam, M.M., Islam, M., Islam, S.M.S., Islami, F., Iso, H., Ivers, R.Q., Iwu, C.C.D., Iyamu, I.O., Jaafari, J., Jacobsen, K.H., Jadidi-Niaragh, F., Jafari, H., Jafarinia, M., Jahagirdar, D., Jahani, M.A., Jahanmehr, N., Jakovljevic, M., Jalali, A., Jalilian, F., James, S.L., Janjani, H., Janodia, M.D., Jayatilleke, A.U., Jeemon, P., Jenabi, E., Jha, R.P., Jha, V., Ji, J.S., Jia, P., John, O., John-Akinola, Y.O., Johnson, C.O., Johnson, S.C., Jonas, J.B., Joo, T., Joshi, A., Jozwiak, J.J., Jürisson, M., Kabir, A., Kabir, Z., Kalani, H., Kalani, R., Kalankesh, L.R., Kalhor, R., Kamiab, Z., Kanchan, T., Karami Matin, B., Karch, A., Karim, M.A., Karimi, S.E., Kassa, G.M., Kassebaum, N.J., Katikireddi, S.V., Kawakami, N., Kayode, G.A., Keddie, S.H., Keller,

C., Kereselidze, M., Khafaie, M.A., Khalid, N., Khan, M., Khatab, K., Khater, M.M., Khatib, M.N., Khayamzadeh, M., Khodayari, M.T., Khundkar, R., Kianipour, N., Kieling, C., Kim, D., Kim, Y.-E., Kim, Y.J., Kimokoti, R.W., Kisa, A., Kisa, S., Kissimova-Skarbek, K., Kivimäki, M., Kneib, C.J., Knudsen, A.K.S., Kocarnik, J.M., Kolola, T., Kopec, J.A., Kosen, S., Koul, P.A., Koyanagi, A., Kravchenko, M.A., Krishan, K., Krohn, K.J., Kuate Defo, B., Kucuk Bicer, B., Kumar, G.A., Kumar, M., Kumar, P., Kumar, V., Kumaresh, G., Kurmi, O.P., Kusuma, D., Kyu, H.H., La Vecchia, C., Lacey, B., Lal, D.K., Lalloo, R., Lam, J.O., Lami, F.H., Landires, I., Lang, J.J., Lansingh, V.C., Larson, S.L., Larsson, A.O., Lasrado, S., Lassi, Z.S., Lau, K.M.-M., Lavados, P.M., Lazarus, J. V., Ledesma, J.R., Lee, P.H., Lee, S.W.H., LeGrand, K.E., Leigh, J., Leonardi, M., Lescinsky, H., Leung, J., Levi, M., Lewington, S., Li, S., Lim, L.-L., Lin, C., Lin, R.-T., Linehan, C., Linn, S., Liu, H.-C., Liu, S., Liu, Z., Looker, K.J., Lopez, A.D., Lopukhov, P.D., Lorkowski, S., Lotufo, P.A., Lucas, T.C.D., Lugo, A., Lunevicius, R., Lyons, R.A., Ma, J., MacLachlan, J.H., Maddison, E.R., Maddison, R., Madotto, F., Mahasha, P.W., Mai, H.T., Majeed, A., Maled, V., Maleki, S., Malekzadeh, R., Malta, D.C., Mamun, A.A., Manafi, A., Manafi, N., Manguerra, H., Mansouri, B., Mansournia, M.A., Mantilla Herrera, A.M., Maravilla, J.C., Marks, A., Martins-Melo, F.R., Martopullo, I., Masoumi, S.Z., Massano, J., Massenburg, B.B., Mathur, M.R., Maulik, P.K., McAlinden, C., McGrath, J.J., McKee, M., Mehndiratta, M.M., Mehri, F., Mehta, K.M., Meitei, W.B., Memiah, P.T.N., Mendoza, W., Menezes, R.G., Mengesha, E.W., Mengesha, M.B., Mereke, A., Meretoja, A., Meretoja, T.J., Mestrovic, T., Miazgowski, B., Miazgowski, T., Michalek, I.M., Mihretie, K.M., Miller, T.R., Mills, E.J., Mirica, A., Mirrakhimov, E.M., Mirzaei, H., Mirzaei, M., Mirzaei-Alavijeh, M., Misganaw, A.T., Mithra, P., Moazen, B., Moghadaszadeh, M., Mohamadi, E., Mohammad, D.K., Mohammad, Y., Mohammad Gholi Mezerji, N., Mohammadian-Hafshejani, A., Mohammadifard, N., Mohammadpourhodki, R., Mohammed, S., Mokdad, A.H., Molokhia, M., Momen, N.C., Monasta, L., Mondello, S., Mooney, M.D., Moosazadeh, M., Moradi, G., Moradi, M., Moradi-Lakeh, M., Moradzadeh, R., Moraga, P., Morales, L., Morawska, L., Moreno Velásquez, I., Morgado-da-Costa, J., Morrison, S.D., Mosser, J.F., Mouodi, S., Mousavi, S.M., Mousavi Khaneghah, A., Mueller, U.O., Munro, S.B., Muriithi, M.K., Musa, K.I., Muthupandian, S., Naderi, M., Nagarajan, A.J., Nagel, G., Naghshtabrizi, B., Nair, S., Nandi, A.K., Nangia, V., Nansseu, J.R., Nayak, V.C., Nazari, J., Negoi, I., Negoi, R.I., Netsere, H.B.N., Ngunjiri, J.W., Nguyen, C.T., Nguyen, J., Nguyen, M., Nguyen, M., Nichols, E., Nigatu, D., Nigatu, Y.T., Nikbakhsh, R., Nixon, M.R., Nnaji, C.A., Nomura, S., Norrving, B., Noubiap, J.J., Nowak, C., Nunez-Samudio, O., Oţoiu, A., Oancea, B., Odell, C.M., Ogbo, F.A., Oh, I.-H., Okunga, E.W., Oladnabi, M., Olagunju, A.T., Olusanya, B.O., Olusanya, J.O., Oluwasanu, M.M., Omar Bali, A., Omer, M.O., Ong, K.L., Onwujekwe, O.E., Orji, A.U., Orpana, H.M., Ortiz, A., Ostroff, S.M., Otstavnov, N., Otstavnov, S.S., Øverland, S., Owolabi, M.O., P A, M., Padubidri, J.R., Pakhare, A.P., Palladino, R., Pana, A., Panda-Jonas, S., Pandey, A., Park, E.-K., Parmar, P.G.K., Pasupula, D.K., Patel, S.K., Paternina-Caicedo, A.J., Pathak, A., Pathak, M., Patten, S.B., Patton, G.C., Paudel, D., Pazoki Toroudi, H., Peden, A.E., Pennini, A., Pepito, V.C.F., Peprah, E.K., Pereira, A., Pereira, D.M., Perico, N., Pham, H.Q., Phillips, M.R., Pigott, D.M., Pilgrim, T., Pilz, T.M., Pirsaheb, M., Plana-Ripoll, O., Plass, D., Pokhrel, K.N., Polibin, R. V, Polinder, S., Polkinghorne, K.R., Postma, M.J., Pourjafar, H., Pourmalek, F., Pourmirza Kalhori, R., Pourshams, A., Poznańska, A., Prada, S.I., Prakash, V., Pribadi, D.R.A., Pupillo, E., Quazi Syed, Z., Rabiee, M., Rabiee, N., Radfar, A., Rafiee, A., Rafiei, A.,

Raggi, A., Rahimi-Movaghar, A., Rahman, M.A., Rajabpour-Sanati, A., Rajati, F., Ramezanzadeh, K., Ranabhat, C.L., Rao, P.C., Rao, S.J., Rasella, D., Rastogi, P., Rathi, P., Rawaf, D.L., Rawaf, S., Rawal, L., Razo, C., Redford, S.B., Reiner, R.C., Reinig, N., Reitsma, M.B., Remuzzi, G., Renjith, V., Renzaho, A.M.N., Resnikoff, S., Rezaei, N., Rezai, M. sadegh, Rezapour, A., Rhinehart, P.-A., Riahi, S.M., Ribeiro, A.L.P., Ribeiro, D.C., Ribeiro, D., Rickard, J., Roberts, N.L.S., Roberts, S., Robinson, S.R., Roever, L., Rolfe, S., Ronfani, L., Roshandel, G., Roth, G.A., Rubagotti, E., Rumisha, S.F., Sabour, S., Sachdev, P.S., Saddik, B., Sadeghi, E., Sadeghi, M., Saeidi, S., Safi, S., Safiri, S., Sagar, R., Sahebkar, A., Sahraian, M.A., Sajadi, S.M., Salahshoor, M.R., Salamati, P., Salehi Zahabi, S., Salem, H., Salem, M.R.R., Salimzadeh, H., Salomon, J.A., Salz, I., Samad, Z., Samy, A.M., Sanabria, J., Santomauro, D.F., Santos, I.S., Santos, J.V., Santric-Milicevic, M.M., Saraswathy, S.Y.I., Sarmiento-Suárez, R., Sarrafzadegan, N., Sartorius, B., Sarveazad, A., Sathian, B., Sathish, T., Sattin, D., Sbarra, A.N., Schaeffer, L.E., Schiavolin, S., Schmidt, M.I., Schutte, A.E., Schwebel, D.C., Schwendicke, F., Senbeta, A.M., Senthilkumaran, S., Sepanlou, S.G., Shackelford, K.A., Shadid, J., Shahabi, S., Shaheen, A.A., Shaikh, M.A., Shalash, A.S., Shams-Beyranvand, M., Shamsizadeh, M., Shannawaz, M., Sharafi, K., Sharara, F., Sheena, B.S., Sheikhtaheri, A., Shetty, R.S., Shibuya, K., Shiferaw, W.S., Shigematsu, M., Shin, J. Il, Shiri, R., Shirkoohi, R., Shrime, M.G., Shuval, K., Siabani, S., Sigfusdottir, I.D., Sigurvinsdottir, R., Silva, J.P., Simpson, K.E., Singh, A., Singh, J.A., Skiadaresi, E., Skou, S.T., Skryabin, V.Y., Sobngwi, E., Sokhan, A., Soltani, S., Sorensen, R.J.D., Soriano, J.B., Sorrie, M.B., Soyiri, I.N., Sreeramareddy, C.T., Stanaway, J.D., Stark, B.A., Ştefan, S.C., Stein, C., Steiner, C., Steiner, T.J., Stokes, M.A., Stovner, L.J., Stubbs, J.L., Sudaryanto, A., Sufiyan, M.B., Sulo, G., Sultan, I., Sykes, B.L., Sylte, D.O., Szócska, M., Tabarés-Seisdedos, R., Tabb, K.M., Tadakamadla, S.K., Taherkhani, A., Tajdini, M., Takahashi, K., Taveira, N., Teagle, W.L., Teame, H., Tehrani-Banihashemi, A., Teklehaimanot, B.F., Terrason, S., Tessema, Z.T., Thankappan, K.R., Thomson, A.M., Tohidinik, H.R., Tonelli, M., Topor-Madry, R., Torre, A.E., Touvier, M., Tovani-Palone, M.R.R., Tran, B.X., Travillian, R., Troeger, C.E., Truelsen, T.C., Tsai, A.C., Tsatsakis, A., Tudor Car, L., Tyrovolas, S., Uddin, R., Ullah, S., Undurraga, E.A., Unnikrishnan, B., Vacante, M., Vakilian, A., Valdez, P.R., Varughese, S., Vasankari, T.J., Vasseghian, Y., Venketasubramanian, N., Violante, F.S., Vlassov, V., Vollset, S.E., Vongpradith, A., Vukovic, A., Vukovic, R., Waheed, Y., Walters, M.K., Wang, J., Wang, Y., Wang, Y.-P., Ward, J.L., Watson, A., Wei, J., Weintraub, R.G., Weiss, D.J., Weiss, J., Westerman, R., Whisnant, J.L., Whiteford, H.A., Wiangkham, T., Wiens, K.E., Wijeratne, T., Wilner, L.B., Wilson, S., Wojtyniak, B., Wolfe, C.D.A., Wool, E.E., Wu, A.-M., Wulf Hanson, S., Wunrow, H.Y., Xu, G., Xu, R., Yadgir, S., Yahyazadeh Jabbari, S.H., Yamagishi, K., Yaminfirooz, M., Yano, Y., Yaya, S., Yazdi-Feyzabadi, V., Yearwood, J.A., Yeheyis, T.Y., Yeshitila, Y.G., Yip, P., Yonemoto, N., Yoon, S.-J., Yoosefi Lebni, J., Younis, M.Z., Younker, T.P., Yousefi, Z., Yousefifard, M., Yousefinezhadi, T., Yousuf, A.Y., Yu, C., Yusefzadeh, H., Zahirian Moghadam, T., Zaki, L., Zaman, S. Bin, Zamani, M., Zamanian, M., Zandian, H., Zangeneh, A., Zastrozhin, M.S., Zewdie, K.A., Zhang, Y., Zhang, Z.-J., Zhao, J.T., Zhao, Y., Zheng, P., Zhou, M., Ziapour, A., Zimsen, S.R.M., Naghavi, M. & Murray, C.J.L. 2020. Global burden of 369 diseases and injuries in 204 countries and territories, 1990–2019: a systematic analysis for the Global Burden of Disease Study 2019. *The Lancet* 396(10258): 1204–1222.

WHO. 2014. Taeniasis / cysticercosis. *Fact sheets*: 4 pages.

Wibawa, T. & Satoto, T.B.T. 2016. Magnitude of Neglected Tropical Diseases in Indonesia at Postmillennium Development Goals Era. *Journal of Tropical Medicine* 2016.

Wiersinga, W.J., Virk, H.S., Torres, A.G., Currie, B.J., Peacock, S.J., Dance, D.A.B. & Limmathurotsakul, D. 2018. Melioidosis. *Nature Reviews Disease Primers* 4.

Winkler, A.S., Klohe, K., Schmidt, V., Haavardsson, I., Abraham, A., Prodjinotho, U.F., Ngowi, B., Sikasunge, C., Noormahomed, E., Amuasi, J., Kaducu, J., Ngowi, H., Abele-Ridder, B., Harrison, W.E. & Prazeres da Costa, C. 2018. Neglected tropical diseases – the present and the future. *Tidsskrift for Den norske legeforening*(born 1971).

Winkler, D.A. 2021. Use of Artificial Intelligence and Machine Learning for Discovery of Drugs for Neglected Tropical Diseases. *Frontiers in Chemistry* 9(March): 1–15.

Wolff, N.S., Jacobs, M.C., Wiersinga, W.J. & Hugenholtz, F. 2021. Pulmonary and intestinal microbiota dynamics during Gram-negative pneumonia-derived sepsis. *Intensive Care Medicine Experimental* 9(1).

Wong, R.R., Baharom, K.M., Ghazali, A.K., Hassan, A.K.R. & Nathan, S. 2021. Phenotype and virulence assessment of a Burkholderia pseudomallei soil isolate from Malaysia. *Sains Malaysiana* 50(5): 1233–1241.

World Health Assembly. 1997. Elimination of lymphatic filariasis as a public health problem (50th World Health Assembly).

World Health Organization. 2018. Dengue vaccine: WHO position paper - September 2018. *Who* 93(36): 457–476.

World Health Organization. 2019. Laboratory techniques in rabies. Fifth edition. Volume 1. *Laboratory techniques in rabies* 1: 304.

World Health Organization. 2020. Global Health Estimates 2020: Disease burden by Cause, Age, Sex, by Country and by Region, 2000-2019.

World Health Organization. 2021. Virtual Meeting of Regional Technical Advisory Group for dengue and other arbovirus diseases(October): 4–6.

World Health Organization. 2022a. Manual for monitoring insecticide resistance in mosquito vectors and selecting appropriate interventions. *Organização Mundial da Saúde*, hlm.

World Health Organization. 2022b. Taenia solium - Use of existing diagnostic tools in public health programmes. *World Health Organization*, hlm.

World Health Organization. 2022c. Global programme to eliminate lymphatic filariasis:

progress report, 2021. *WHO Weekly Epidemiological Record*, hlm.

World Organisation for Animal Health. 2008. Rabies Technical Disease Information: 1–4.

Yajima, A. & Ichimori, K. 2021. Progress in the elimination of lymphatic filariasis in the Western Pacific Region: Successes and challenges. *International Health* 13: S10–S16.

You, Y., Lai, X., Pan, Y., Zheng, H., Vera, J., Liu, S., Deng, S. & Zhang, L. 2022. Artificial intelligence in cancer target identification and drug discovery. *Signal Transduction and Targeted Therapy* 7(1): 1–24.

Zeynudin, A., Degefa, T., Tesfaye, M., Suleman, S., Yesuf, E.A., Hajikelil, Z., Ali, S., Azam, K., Husen, A., Yasin, J. & Wieser, A. 2022. Prevalence and intensity of soil-Transmitted helminth infections and associated risk factors among household heads living in the peri-urban areas of Jimma town, Oromia, Ethiopia: A community-based cross-sectional study. *PLoS ONE* 17(9 September): 1–17.

Zhou, G., Chen, M., Ju, C.J.T., Wang, Z., Jiang, J.Y. & Wang, W. 2020. Mutation effect estimation on protein–protein interactions using deep contextualized representation learning. *NAR Genomics and Bioinformatics* 2(2): 1–12.

Zhu, H.H., Huang, J.L., Zhu, T.J., Zhou, C.H., Qian, M.B., Chen, Y.D. & Zhou, X.N. 2020. National surveillance on soil-transmitted helminthiasis in the People's Republic of China. *Acta Tropica* 205(January): 105351.